\documentclass[aps,twocolumn,floats,longbibliography,superscriptaddress]{revtex4-2}
\usepackage{graphics,graphicx,epsfig}
\usepackage{amssymb,color}
\usepackage{epsf,epstopdf}
\usepackage {amsmath}
\usepackage{ragged2e}
\usepackage[compact]{titlesec}

\usepackage[final,hidelinks]{hyperref}

\newcommand{\beq}{\begin{equation}}
\newcommand{\eeq}{\end{equation}}
\newcommand{\beqn}{\begin{eqnarray}}
\newcommand{\eeqn}{\end{eqnarray}}

\titlespacing{\section}{0pt}{5pt plus 4pt minus 2pt}{5pt plus 2pt minus 2pt}
\begin{document}

\setlength{\parskip}{6pt}

\title{Precise and scalable self-organization in mammalian pseudo-embryos}

\author{M\'elody Merle}
\thanks{These authors contributed equally.}
\affiliation{Department of Developmental and Stem Cell Biology, CNRS UMR3738 Paris Cit\'e, Institut Pasteur, Paris, France}
\author{Leah Friedman}
\thanks{These authors contributed equally.}
\affiliation{Department of Developmental and Stem Cell Biology, CNRS UMR3738 Paris Cit\'e, Institut Pasteur, Paris, France}
\author{Corinne Chureau}
\thanks{These authors contributed equally.}
\affiliation{Department of Developmental and Stem Cell Biology, CNRS UMR3738 Paris Cit\'e, Institut Pasteur, Paris, France}
\author{Armin Shoushtarizadeh}
\affiliation{Department of Developmental and Stem Cell Biology, CNRS UMR3738 Paris Cit\'e, Institut Pasteur, Paris, France}
\author{Thomas Gregor}
\email[Correspondence: ]{thomas.gregor@pasteur.fr}
\affiliation{Department of Developmental and Stem Cell Biology, CNRS UMR3738 Paris Cit\'e, Institut Pasteur, Paris, France}
\affiliation{Joseph Henry Laboratories of Physics \& Lewis-Sigler Institute for Integrative Genomics, Princeton University, Princeton, NJ, USA}

\date{\today}

\begin{abstract}
Gene expression is inherently noisy, posing a challenge to understanding how precise and reproducible patterns of gene expression emerge in mammals. We investigate this phenomenon using gastruloids, an \textit{in vitro} model for early mammalian development. Our study reveals intrinsic reproducibility in the self-organization of gastruloids, encompassing growth dynamics and gene expression patterns. We observe a remarkable degree of control over gene expression along the main body axis, with pattern boundaries positioned at single-cell precision. Furthermore, as gastruloids grow, both their physical proportions and gene expression patterns scale proportionally with system size. Notably, these properties emerge spontaneously in self-organizing cell aggregates, distinct from many \textit{in vivo} systems constrained by fixed boundary conditions. 
Our findings shed light on the intricacies of developmental precision, reproducibility, and size scaling within a mammalian system, suggesting that these phenomena might constitute fundamental features of multicellularity.
\end{abstract}
\maketitle

\noindent Multicellular development entails the meticulous organization of cellular identities and body proportions in both spatial and temporal dimensions~\cite{Conklin1905,Kirschner1997,Houchmandzadeh2002}. Gastrulation is a key event in this process, during which the body plan and the subsequent establishment of asymmetric body axes occur. Coordinated gene expression during this stage leads to reproducible patterns between individuals despite the noisiness of the underlying molecular events of gene regulation~\cite{Arias2006,Briscoe2015}. 

The challenge of translating transcriptional variability into precise and reproducible gene expression patterns has captivated research across a spectrum of animal models, from nematodes to vertebrates~\cite{Sulston1983, Bollenbach2008, Bier2015, Bentovim2017, Zagorski2017, Guignard2020}. Developmental processes have been conceptualized as a sequence of steps aimed at mitigating and correcting errors in the face of molecular noise~\cite{Waddington1942,Arias2006}. In vertebrates, mechanisms such as differential specification rates and cell sorting have been described as error-correction strategies
~\cite{Kicheva2014, Tsai2020}.

However, in the context of the early fly embryo, the precision of macroscopic body plan features can be traced back to the precision of maternal inputs~\cite{Petkova2014}. Exemplified by the morphogen gradient of Bicoid~\cite{Driever1988}, this precision is transmitted at the single-cell level along the major body axis to zygotic genes before gastrulation~\cite{Petkova2019}. Such precision in flies challenges the limits of molecular noise~\cite{Gregor2007,Dubuis2013a}, suggesting that successive developmental stages may have evolved to minimize noise transmission at each step, both across evolutionary time scales and within the spatiotemporal boundaries of individual organisms~\cite{Lacalli2022}.

An intriguing consequence of this precision is the scaling of gene expression patterns relative to system size~\cite{Houchmandzadeh2002,Antonetti2018}. Scaling, observed in both invertebrates and vertebrates~\cite{Ishimatsu2018, Uygur2016, Almuedo2018, Leibovich2020}, entails the preservation of body plan proportions among different individuals. During development, scaling manifests at various levels, encompassing morphogenetic movements, gene expression domains, and other phenomena, reflecting the intricate interplay of regulatory mechanisms~\cite{ Al2020, Cheung2011, Ben2011, Huang2020, Almuedo2018, Romanova2022, Saiz2020}.

In contrast to organisms with well-defined developmental boundaries, such as flies, frogs, or worms, mammalian development relies on self-organization and continuous growth. Quantitative assessments of reproducibility, precision, and scaling in mammalian systems have been limited, prompting inquiries into whether the precision observed in flies is even necessary, as mammals rely on different developmental mechanisms. These properties have been found in other types of self-organizing systems, such as scaling in flatworms during regeneration~\cite{Stuckemann2017}, suggesting scaling could also be achieved during self-organization-driven development.

Recent progress with \textit{in vitro} models \cite{Gritti2021,Rosado2021}, including gastruloids~\cite{VandenBrink2014}, derived from mouse embryonic stem cells (mESCs), present promising avenues for investigation. These three-dimensional pseudo-embryos mimic critical events of mammalian gastrulation through self-organized patterning. They break symmetry and elongate along an axis that resembles the most posterior part of the mouse embryo's anterior-posterior (AP) axis~\cite{Beccari2018} can be cultivated in substantial quantities, rendering them conducive to quantitative approaches~\cite{Hashmi2022,Underhill2023}. Yet, concerns regarding the reproducibility of these systems have been raised~\cite{Fu2021}.

\begin{figure*}
\centering
\includegraphics[width=0.78\textwidth]{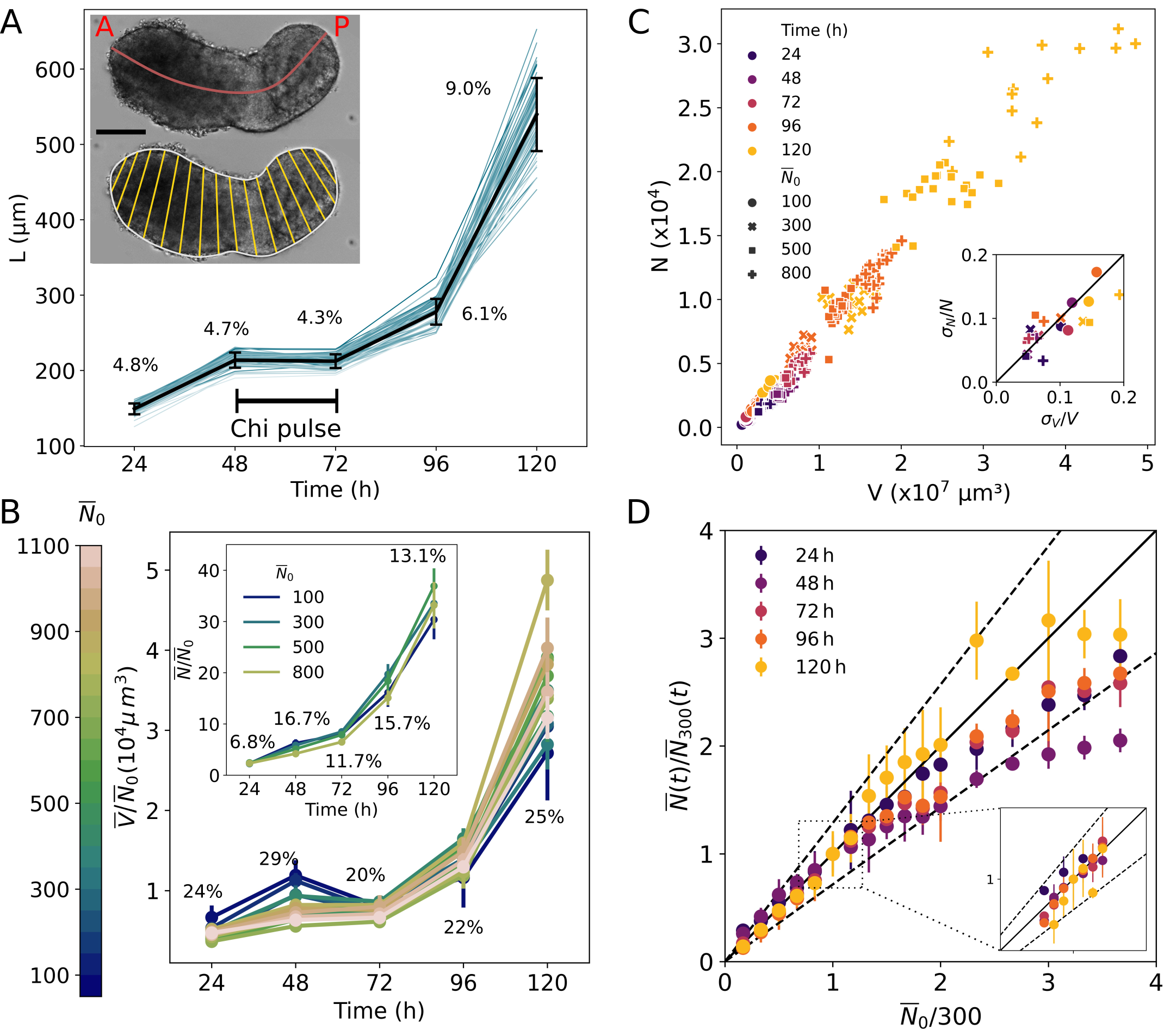} 
\caption{Reproducible gastruloid growth, scaled to system size. 
{\bf A}:~Gastruloid midline length variation as a function of time. Curves shown for 57 gastruloids followed individually over time (blue) and mean (black). Percent variation around the mean is reported for each time point. Spread of initial number of seeded mESCs is $N_0=305\pm28$ cells (Fig.~S1A). (For an equivalent relationship for volume see Fig.~S2A.) Inset shows a brightfield image of a gastruloid at $120\,{\rm h}$, overlaid with its midline ranging from anterior (A) to posterior (P) pole (red, top) and sliced evenly for volume reconstruction (yellow, bottom); scalebar is  $100\,$µm; also see Fig.~S1B-D.  
{\bf B}:~Gastruloid volume and total cell count (inset and Fig.~S1E) as a function of time. Volumes are normalized by the average number of initial seed cells $\overline{N}_0$ at time zero (color code). Each line represents the mean of on average 15 gastruloids with the same $N_0$. Percentages correspond to residual variations within which normalized volumes collapse for 17 different values of $\overline{N}_0$. Similar collapse for normalized gastruloid cell counts for four values of $\overline{N}_0$ (inset).  
{\bf C}:~Scatter plot of total cell count versus the measured volume for 492 individual gastruloids at different time points (color code) and with varying $\overline{N}_0$ (symbol); Pearson correlation coefficient is $r=0.99$. Inset shows correlation ($r=0.78$) of variability for $N$ and $V$ for sets of gastruloids with identical age and $N_0$. This is evidence that the independent methods for measuring $N$ and $V$ are accurate estimates of gastruloid growth.  
{\bf D}:~Cell count $\overline{N}(t)/\overline{N}_{300}(t)$ as a function of the initial seed cell count $\overline{N}_0/300$ in units of the reference seed at $\overline{N}_0=300$. Time is encoded by color (see legend). Black diagonal (slope = 1) represents perfect scaling of gastruloid size at time $t$ upon changes in $\overline{N}_0$ ranging over $50\le \overline{N}_0 \le 1100$. Dashed line estimates expected deviations from perfect scaling at $120\,{\rm h}$ due to fluctuations in $\overline{N}_0/300$ and in the doubling time $t_{\rm D}$ given a simple exponential growth model (Fig.~S2E and Methods). Detailed representations for individual time points can be found in Fig.~S2G. Inset shows the same relationship centered around $\overline{N}_0=300$ where the regression slope is statistically indistinguishable from one at all time points (see Table~S1).
}
\label{fig1}
\end{figure*}


\section*{RESULTS}

\noindent{\bf Reproducible gastruloid growth and size scaling.} Quantitative analysis in mammalian development faces a significant hurdle due to the inherent predicament of achieving experiment reproducibility, particularly when replicating experiments under identical conditions is challenging. The protracted and often inaccessible nature of embryos further complicates this issue. However, the gastruloid model presents a promising solution to these challenges, offering the capability to culture hundreds of specimens simultaneously and under equivalent conditions.

Under tightly controlled experimental conditions, we have maximized the degree of reproducibility in gastruloid cultures. Within the confines of the original protocol~\cite{Beccari2018b}, we achieved a 97\% success rate in inducing the elongation of gastruloids along a single axis (Fig.~S1). To further assess the intrinsic reproducibility of the self-organization processes within these systems, we examined general physical properties, including the uniformity of growth and the influence of initial conditions. We quantified growth dynamics by monitoring the length of the midline, the total volume, and the total cell count of individual gastruloids over a five-day period (Fig.~S1, Methods). 

\begin{figure*}
\centering
\includegraphics[width=\textwidth]{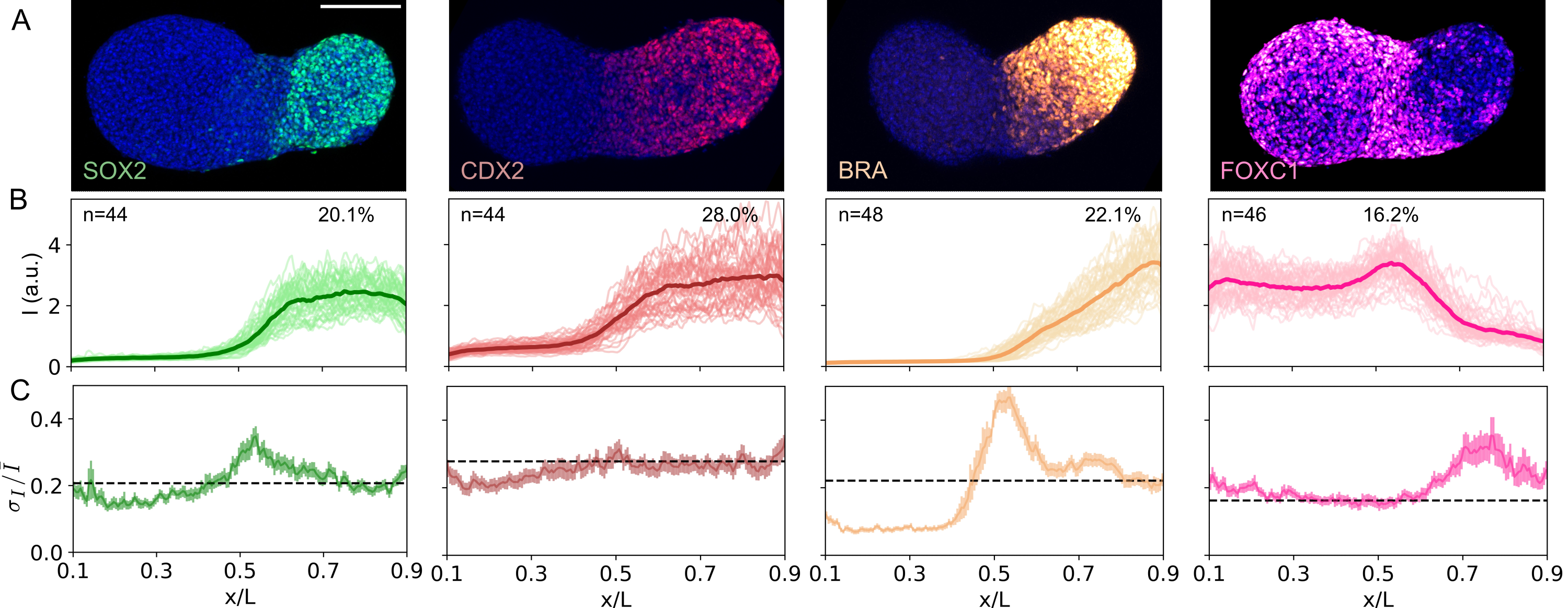} 
\caption{Reproducibility of gene expression patterns in gastruloids. 
{\bf A}:~Maximum projections of four confocal image stacks of  $120\,{\rm h}$ old gastruloids stained by immunofluorescence for SOX2, CDX2, BRA, FOXC1. AP-axis is in a left--right orientation. Scalebar is $100\,$µm. 
{\bf B}:~n individual raw gene expression profiles (light color) for the four markers in A and the corresponding average profile (dark bold) projected on the midline and reported relative to gastruloid length $L$. 
{\bf C}:~Variability ($\sigma_I/\overline{I}$) of the respective gene expression patterns from B as a function of relative position along the midline $x/L$. Error bars are obtained by bootstrapping. Dashed lines represent the average variability in the region where genes are most highly expressed (values in B, see Fig.~S5B).} 
\end{figure*}

The growth curves of individual gastruloids exhibit a remarkable degree of consistency, as evidenced by the convergence of both length and volume measurements over time (Fig.~1A and S2A). This observation underscores the high reproducibility of growth dynamics at various time points. The residual variability observed in these growth curves can, in part, be attributed to two primary factors. First, it is influenced by the variability in the initial number of seeded cells $N_0$ (Fig.~S1A). Additionally, fluctuations in the effective doubling time, which we have measured to average at $26.4\pm 1.7\,$h for $\overline{N}_0=300$ (Fig.~S2B), contribute to this variability.

The volumes of gastruloids display a significant correlation with the initial number of seeded cells ($N_0$) at all measured time points (Fig.~S2C). Moreover, when we extend our investigation to encompass a wide range of average $\overline{N}_0$ values (up to 22-fold changes), these correlations become significantly more pronounced (Fig.~S2D). It is worth noting that as the average $\overline{N}_0$ increases, the percentage of gastruloids exhibiting multipolarity also increases (Fig.~S1D). Yet, our volume measurement algorithm is not applicable to multipolar gastruloids, introducing a bias in the data, which becomes more apparent as $\overline{N}_0$ increases. Nevertheless, a significant finding emerges from our data: growth curves of gastruloid volumes, when normalized by $\overline{N}_0$, consistently converge (Fig.~1B). This convergence serves as compelling evidence that growth is not only reproducible across all observed time points but also scales in relation to the initial number of seeded mESCs.

A similar relationship becomes apparent when examining the growth curves of the total number of cells, $N$ (obtained from chemically dissociated gastruloids, Fig.~S1E). For varying initial $\overline{N}_0$ values, the growth curves converge, as illustrated in Fig.~1B (inset), while the effective doubling time remains consistent (Fig.~S2E). The residual spread in these converging growth curves can be attributed to other factors, such as variability between experiments (Fig.~S2F).

The observed common relationship between growth and developmental time hints at the emergence of size control as an intrinsic property within gastruloids. Interestingly, it suggests that a refinement mechanism aimed at achieving specific sizes with error reduction may not be a necessary component of this system. Instead, the scaling of gastruloid size with the initial seed number implies a departure from the principles governing mouse embryos~\cite{Snow1979,Lewis1982,Rands1986}.
Unlike the self-regulation mechanisms observed in mouse embryos, our findings suggest that gastruloid growth dynamics operate independently of system size, with absolute size (both in terms of cell number and volume) being directly linked to the initial seeding number.

To substantiate this assertion, we conducted a direct assessment of the extent to which gastruloid volume can serve as a predictor of the total cell count. Measurements of cell count and volume within the same gastruloid revealed a strong, linear relationship that remained consistent across various time points and different average $\overline{N}_0$ values (Fig.~1C). This strong correlation suggests that the inherent dispersion in cell size is remarkably conserved across diverse gastruloids and under varying external conditions (Fig.~S8G). 

To further validate our observations, we compared our results (Fig.~1B) with the hypothesis of perfect scaling. Perfect scaling, in this context, denotes a linear relationship between $N(t)$ and $N_0$. When we represent these values with respect to the reference seeding number $\overline{N}_0=300$ (Fig.~1D), perfect scaling is achieved by a slope$\,=1$ (black line). It signifies that starting with twice as many cells (in units of $\overline{N}_0=300$) results in precisely twice as many cells at any given time point. We utilized the measured errors associated with both the initial seeding number and the doubling time to estimate the expected error at 120 hours.

For nearly all stages of gastruloid growth, the data points fall within the boundaries of expected deviations. Notably, at earlier time points ($24$h and $48$h), growth scaling is observed in a range of $\overline{N}_0$ values spanning between $100$ and $600$ (Fig.~S2G). In the case $\overline{N}_0=300$ (Fig.~1D, inset), the slope is statistically indistinguishable from one at all time points. These results collectively underscore the high fidelity of the monitored 5-day growth process. Under carefully controlled experimental conditions, gastruloids exhibit a remarkable capacity for self-organization with meticulous control over the variability in growth rate (Fig.~S2E) and other noisy processes of size regulation.

\vspace{.2 cm}
\noindent{\bf Reproducible gene expression patterning.} 
Coordinated growth and, in particular, axis elongation in gastruloids are closely associated with the expression patterns of genes along the body axes. Consequently, we ask how the observed physical properties of reproducibility and scaling manifest in the anterior-posterior (AP) patterning of gene expression. To this end, we conducted gene expression profile measurements five days after seeding, a time point when the pseudo-AP axis is morphologically well-established. Our focus was on the AP patterning of four germ-layer markers, namely BRA, CDX2, FOXC1, and SOX2. These markers are well-documented for their pivotal roles in the differentiation of tissue progenitors and the establishment of the AP axis during gastrulation~\cite{Mittnenzweig2021,Neijts2014,Amin2016,Blassberg2022}.

We performed immunofluorescence staining (Fig.~2A), and from two-dimensional (2D) maximum projections of a confocal image stack, we extracted one-dimensional (1D) intensity profiles projected isometrically along each gastruloid's midline (Fig.~2B, Fig.~S3B, Methods). Unexpectedly, individual profiles for all four genes in mammalian systems are closely clustered around the average profile. The overall variability in these profiles is notably small (Fig.~2C), peaking in regions where gene expression levels change sharply over short distance intervals, such as boundaries between high and low expression domains. 

In pioneering experiments with engineered systems, it has been shown that even when gene expression is highly induced, the resulting expression levels fluctuate~\cite{Elowitz2002, Raser2004}. We observe something similar in the regions where our examined genes are expressed at their maximal levels (Fig.~S6B), with the variability hovering around 20\% (Fig.~2C). This variability near the maximum expression level has also been observed in other organisms~\cite{Gregor2007, Carolina2021}. 

The observed noise may, in principle, result from measurement errors. In separate experiments (see Methods and Fig.~S3, S4, S5, S7, S9), we estimate the component of measurement noise arising in the experimental process. Some sources of experimental noise are inherent to the immunofluorescence staining and imaging processes~\cite{Dubuis2013} (Fig.~S4), while others result from arbitrary choices made during image analysis routines, such as axis definition and the projection method for measuring gene expression patterns (Fig.~S3). Overall, we estimated that all sources of measurement error combined correspond to less than 10\% of the total variance (Fig.~S5, Methods). The values reported here thus represent an upper bound for the biological variability of the system and the true value is even lower (Fig.~S6C). 

\begin{figure}
\centering
\includegraphics[width=\columnwidth]{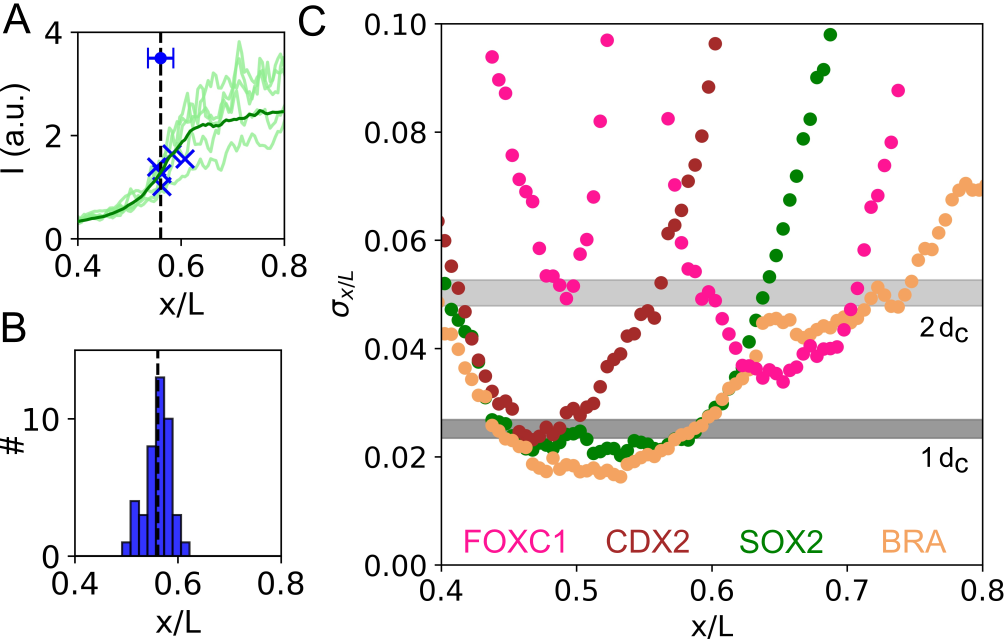}
\caption{Single-cell pattern boundary precision in gastruloids. 
{\bf A}:~Close-up of Fig.~2B for five SOX2 expression profiles as a function of position along the midline (green), with pattern boundary positions of individual profiles marked at the half-maximal expression value (EC50, blue crosses). Mean profile of gastruloid midlines in dark green (n=44). Dashed line is at mean position for these five profiles.
{\bf B}:~Distribution of SOX2 pattern boundary positions from Fig.~2B. The mean defines the pattern boundary position $x_B$ (n=44); the standard deviation of this distribution (blue bar in A) of $2.4\%$ defines the positional error for pattern boundary establishment. 
{\bf C}:~Positional error directly calculated from the standard deviation of intensity values across the individual expression profiles in A, $\sigma_I(x/L)$. For each position $x/L$, this expression error is propagated into an error in position, $\sigma_{x/L}$ (see Methods). Color code as in Fig.~2; gray areas correspond to one and two effective cell diameters $d_c$, respectively, including measurement errors (Fig.~S7 and S8).}
\label{fig3}
\end{figure}

\vspace{.2 cm}
\noindent{\bf Single-cell precision of pattern boundaries.} 
During development, cells rely on patterning signals executed by genes like those analyzed above~\cite{Neijts2014,Amin2016}. However, inherent variability between individual gastruloids (Fig.~2B) limits the precision with which cells can execute their functions and fates at specific positions. We estimate the positional precision for the four analyzed genes by determining the positions along the midline where the half-maximal expression level is reached within the boundary regions for each patterning gene (Fig.~3A and S5A). For instance, in the case of SOX2, we observe a narrow distribution of these positions (Fig.~3B), with a standard deviation of only 2.4\%. The other genes exhibit a similar level of boundary precision (Fig.~S6D). 
 
Instead of focusing solely on a single boundary point, a more comprehensive approach involves considering the entire extent of the pattern and translating the fluctuations in expression levels (Fig.~2C) into positional errors (Fig.~3C, Methods)~\cite{Dubuis2013a}. This broader analysis reveals that a positional precision of 2--4\% is achieved within domains spanning approximately 5--10\% of the gastruloid length. These domains align with the respective boundary regions for each gene (Fig.~S6E). The values obtained through both methods are consistent at the mean pattern boundary positions (Fig.~S6F). In principle, this precision allows cells to use the expression levels of these genes to precisely determine their positions along the pattern boundary.

\begin{figure*}
    \centering
\includegraphics[width=0.95\textwidth]{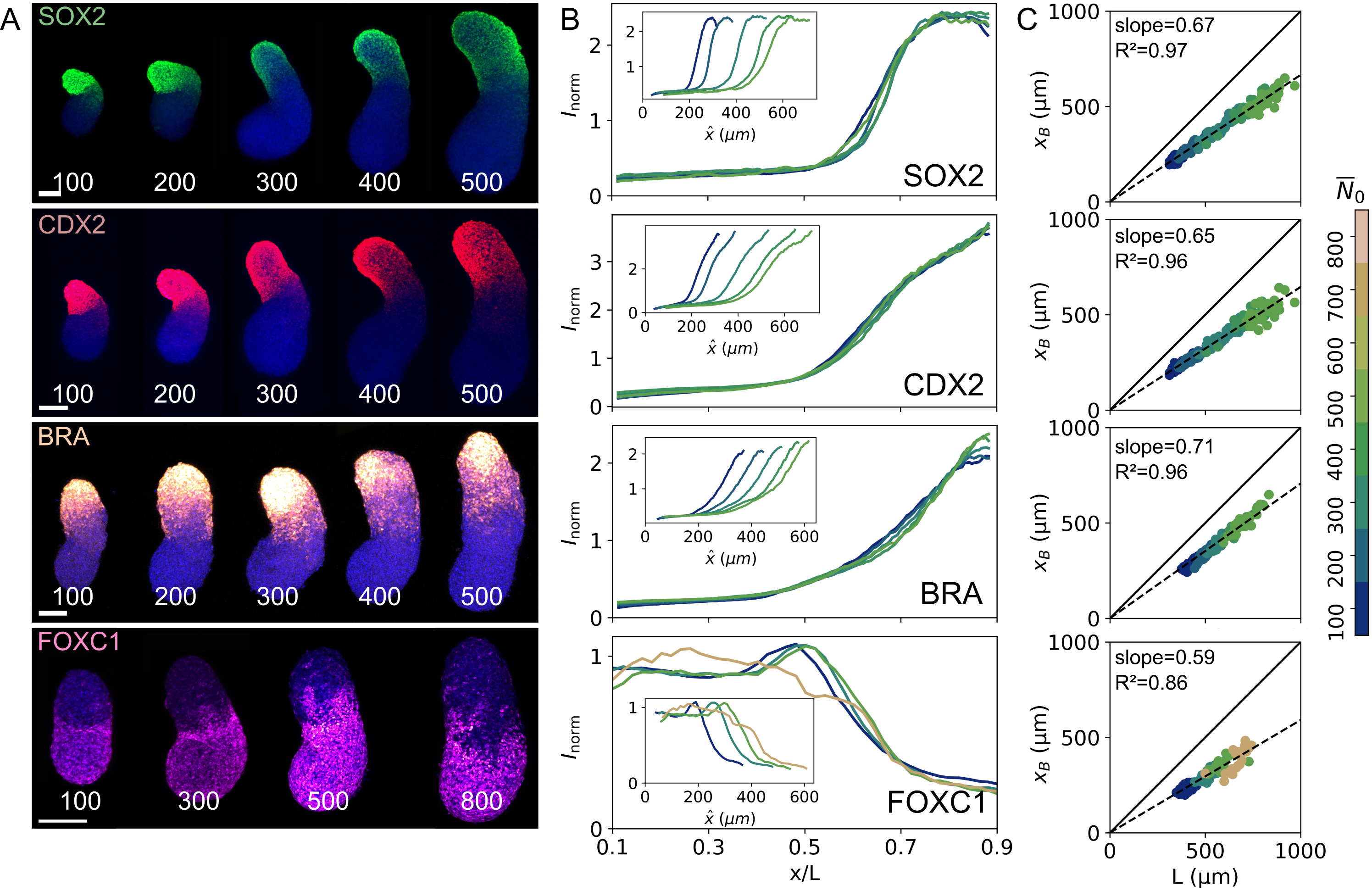}
\caption{Scaling of AP gene expression patterns in gastruloids. 
{\bf A}:~Confocal images of gastruloids immunofluorescently stained for four different genes each representing a different initial seed number $\overline{N}_0$ (in white, also Fig.~S10). AP-axis from bottom to top. Scalebars are $100\,$µm.
{\bf B}:~Normalized mean expression profiles for sets of gastruloids with the same $\overline{N}_0$ (color code in C) as a function of the relative position $x/L$ along the average midline of the respective set (n=15--50 gastruloids per $\overline{N}_0$). AP-axis is in a left--right orientation. Inset shows normalized mean expression profiles as in B as a function of average position in absolute units of the respective set.
{\bf C}:~Boundary position $x_{B}$ in absolute units of individual gastruloids seeded with varying $\overline{N}_0$ (same color code as in B) as a function of absolute individual gastruloid length $L$. Bold diagonal line indicates gastruloid length ($x_{B} = L$). Dashed line shows linear fit with intercept at zero. Perfect scaling would imply $R^{2}=1$, meaning that 100\% of the observed boundary position variance is related to gastruloid length. Here, for the genes SOX2, CDX2, and BRA, the scaling relationship with respect to gastruloid length explains 96--97\% of the boundary position variance.
}
\label{fig4}
\end{figure*}

To gain insight into the practical significance of achieving 2--3\% spatial precision along the midline, we measured the average size of individual cells within gastruloids. We revisited our simultaneous measurement of cell count $N$ and volume $V$ for several hundred gastruloids (Fig.1C). The strong linear relationship $N=s*V$, with a slope $s$ representing the inverse of the mean cell volume, allowed us to determine the effective diameter of cells in developing gastruloids. To validate this measurement, we employed high-resolution 3D reconstructions of individual gastruloids~\cite{Stringer2021}, in which we fluorescently labeled all cell membranes (Fig.~S8). The consensus between both methods yielded an effective cell diameter of $d_c=13.5\pm0.8\,$µm (after 72 hours of development). This value serves as the relevant linear size unit for the system. 

With this system-intrinsic length scale measurement, we determined that the achieved patterning precision corresponds to 1--2 cell diameters along the midline of the gastruloid (Fig.~3C). This finding demonstrates that mammalian gastruloids exhibit patterning precision on par with patterning systems in fly embryos~\cite{Gregor2007}, worms~\cite{Moore2013}, and ascidians~\cite{Guignard2020}. In all these systems, the positional error resulting from gene expression fluctuations allows for distinguishing neighboring cells.

These levels of reproducibility and precision remain consistent for gastruloids grown in parallel from the same population of cells (Fig.~S9). By minimizing sources of variability throughout the entire chain of experimental protocols, from gastruloid seeding to imaging, we obtained profiles with very similar average absolute concentration levels, variability, and positional error in multiple experiments and for various genes. Note that we are reporting the total variance, which includes measurement errors. Therefore, the actual values for the reproducibility and precision of the system are even higher than what we report here. These results provide an absolute scale for the reproducibility of the self-organized patterning process; cells at each point along the pattern produce the same amount of gene product in absolute units. These units translate along the pattern axis into a spatial precision equivalent to the linear dimension of a single cell, arguably the maximum precision that is functionally beneficial for a multicellular system.

\vspace{.2 cm}
\noindent{\bf Gene expression pattern scaling with size.} 
To comprehend the relationship between the growth dynamics of these self-organizing structures and their gene expression profiles, we examine how gene expression patterns scale with gastruloid size. This will shed light on the system's capacity to sustain proportional patterns as it undergoes growth. Despite variations in gastruloid lengths by 7--11\% five days after seeding (Fig.~1A and S2F), the relative positional error in gene expression boundaries remains below 3\%. This suggests that the mechanisms underlying pattern formation in gastruloids can adapt to system size~\cite {Werner2015}. To directly test this hypothesis, we deliberately manipulated the sizes of gastruloids by adjusting the initial number of seeded cells. Within the range where elongation results are most robust, we achieve up to a 2.3-fold change in gastruloid lengths (Fig.~4A and S10). 

For each of the four genes studied earlier, we create sets of immunofluorescently labeled gastruloids with different $\overline{N}_0$. Upon plotting the average gene expression profiles for each $\overline{N}_0$ set against absolute units, they exhibit dispersion along the x-axis in direct proportion to the corresponding average gastruloid length (Fig.~4B, inset). However, upon normalization by the mean gastruloid length within a given set, the average profiles collapse (Fig.~4B), which can also be seen in individual profiles (Fig.~S10C-D). These findings indicate a linear relationship with a zero intercept between the absolute boundary position (as defined in Fig.~3A) and the length in absolute units (Fig.~4C). Our ability to consistently cultivate gastruloids with varying initial numbers of seeded cells $N_0$ enables us to evaluate this scaling relationship across a broad range of gastruloid lengths spanning approximately $600\,$µm.

We can further quantify the scaling effect by examining the position of other key points along the gene expression profiles, such as positions where the intensity equals 25\% and 75\% of the maximum expression (Fig.~S11). We observe that our smallest ($\sim$$300\,$µm) and largest ($\sim$$900\,$µm) gastruloids display boundary shifts of one cell diameter compared to the $\overline{N}_0=300$ reference case, with the exception of BRA, where the shift amounts to 4 cell diameters. Therefore, at each relative position along the gastruloid's midline, a cell consistently produces and maintains an absolute amount of protein with an accuracy within a few tens of percent of its mean value. These results indicate that the expression patterns of these four genes contain information locally about the overall length of the entire system. Furthermore, the residual positional error after rescaling to relative coordinates is consistently within 1--2 cell diameters (Fig.~S11D), for gastruloids with $100\le\overline{N}_0\le500$. Note that this falls within the range for which we observed growth scaling. This level of precision matches the inherent precision of the pattern boundaries, showcasing a remarkable sensitivity to global parameters.


\section*{DISCUSSION}

Our results reveal intriguing properties of mammalian cell aggregates and their potential implications for developmental biology and regenerative medicine. They underline the intrinsic potential for reproducible, precise, and scalable self-organization in gastruloids. These properties seem to go beyond the development of organisms like worms, flies, or frogs and are also achieved by mammalian systems, shedding light on the existence of a general principle of pattern formation that acts at the multicellular scale independently of boundary conditions \cite{Turing1990}.

The remarkable reproducibility, precision, and scalability observed in gastruloids carry profound biological implications. They challenge our conventional understanding of mammalian development, hinting at the precise regulation of developmental features, such as gene expression patterns, during self-organized processes. Additionally, these findings suggest that reproducibility and scaling, evident in both developing embryos and synthetic structures like gastruloids, may represent context-independent properties \cite{Endy2005, Stanton2014}. Collectively, these insights point towards fundamental principles that govern self-organization processes in multicellular systems. 

These properties underscore the importance of exploring how these principles relate to the complex and dynamic processes of \textit{in vivo} development in mammals. Our findings raise questions about the extent to which these principles govern embryonic development and tissue formation. The fact that synthetic systems exhibit intrinsic reproducibility and precision similar to \textit{in vivo} systems expands the possibilities for advanced engineering applications in the field of organoids and, more broadly, cell aggregates \cite{Clevers2016, Fatehullah2016, Lancaster2014, Shariati2021}.

In mammals, achieving these findings is currently not feasible \textit{in vivo}; only \textit{in vitro} systems offer the necessary experimental accessibility and manipulability. Utilizing an \textit{in vitro} system not only provides essential experimental access but also enables precise control over parameters, including system size, and facilitates the ability to continuously perturb the system beyond its natural limits. Our approach and findings suggest that the gastruloid model, and possibly other stem-cell-derived aggregates, have the potential to serve as a powerful tool for quantitative studies of mammalian development and various other biological processes.

It is important to note that while our findings highlight the surprising properties of mammalian cell aggregates, they do not necessarily apply to all features of these systems. For instance, gastruloid shape is not consistently reproducible (as evident in Fig.~2A and 4A). Gastruloid length, which is a component of shape, exhibits more variation at 120 hours compared to pattern boundary positions. This distinction separates self-organized processes, like the emergence of gene expression patterns in these cell aggregates, from those that are predominantly influenced by external and environmental conditions and, therefore, not subject to stringent control~\cite{Kicheva2014,Veenvliet2021}. Only the former appear to be subject to precise regulation.

While the study highlights the potential of gastruloids as a model system, we acknowledge that it has limitations. Gastruloids, while powerful tools, are not identical to \textit{in vivo} embryos, and thus there are aspects of development they cannot fully replicate. Additionally, translating these findings to \textit{in vivo} contexts remains a challenge \cite{Rossi2018,Jensen2023}. Nonetheless, the study underscores the value of using \textit{in vitro} systems as accessible and controllable models for studying development.

In conclusion, our findings provide a fresh perspective on the properties of self-organization in mammalian cell aggregates. These properties are not confined to invertebrate model systems like flies or worms but could be context-independent, spanning millions of years of evolutionary change. Understanding these principles can transform our approach to developmental biology, advancing our knowledge of embryonic development in mammals. Gastruloids and similar \textit{in vitro} systems, with their accessibility and controllability, are poised to play a pivotal role in future research in this field, offering opportunities for quantitative exploration and applications.

\section*{Data and code availability}
Processed immunofluorescence staining data is available as maximum projection images for individual gastruloids, organized by figure number. All images have been deposited on the Zenodo repository under doi: \href{https://zenodo.org/doi/10.5281/zenodo.8108187}{10.5281/zenodo.8108188}. Raw images are available upon request. Custom Python-based analysis code for data processing is available at the GitLab repository (\url{https://gitlab.pasteur.fr/tglab/gastruloids_precisionandscaling}).
 
\section*{Acknowledgments}
We thank Isma Bennabi, David Br\"uckner, Michele Cerminara, Michel Cohen-Tannoudji, Pauline Hansen, Milo\v s Nikoli\'c, Camil Mirdas, Judith Pineau, Jerome Wong-Ng, Benjamin Zoller, and the late Roel Neijts. This work was supported by Institut Pasteur (particularly the cytometry platform), Centre National de la Recherche Scientifique, CFM Foundation for Research, and the French National Research Agency (ANR-10-LABX-73 'Revive', ANR-19-CE45-0016 'Polychrome', ANR-20-CE12-0028 'ChroDynE', and ANR-23-CE13-0021 'GastruCyp').

\section*{Author contributions} 
M.M., L.F., C.C., A.S., and T.G. designed experiments. M.M., L.F., C.C., and A.S. developed experimental protocols. M.M., L.F., and C.C. performed experiments. M.M. and L.F. performed computational image analysis. M.M., L.F., and T.G. wrote the manuscript. T.G. secured funding and supervised the work.

\bibliography{scaling}

\onecolumngrid

\begin{center}
    {\bf APPENDIX}
\end{center}

\appendix
\section{Methods}

\subsection{Cell culture}

129/SvEv mouse embryonic stem cells (mESCs) were cultured on gelatinized tissue-culture 6-wells plates in a humidified incubator (5\% $\rm{CO_2}$, 37°C). They were cultured in Dulbecco’s Modified Eagle’s Medium (DMEM~1X~+~GlutaMAX, Fisher 11584516) supplemented with 10\% fetal bovine serum (FBS, Pansera), 1X Non-Essential Amino Acids (NEAA, Gibco 11140-035), 1$\,$mM Sodium Pyruvate (Gibco 11360-039), 1X Penicillin-Streptomycin (Sigma-Aldrich P4333), 100$\,$µM 2-Mercaptoethanol (Gibco 31350-010), 10$\,$ng/mL leukemia inhibitory factor (LIF, Miltenyi Biotec 130-099-895), 3$\,$µM GSK3 inhibitor CHIR 99021 (Sigma-Aldrich SML1046) and 1$\,$µM MEK inhibitor PDO35901 (Sigma-Aldrich PZ0162). Cells were passaged every other day and seeded at  $5.10^4$$\,$cells/mL using an automatic cell counter (Logos Biosystems LUNA-II$^{\rm TM}$). Media was half-replaced when cells were not passaged. Cells were tested regularly for mycoplasm (Eurofins Mycoplasmacheck).

\subsection{Gastruloid culture}

A complete description of the protocol to generate gastruloids is described in \cite{Beccari2018a}. N2B27 medium was prepared in-house, at least every three weeks. Initial cell seeding was performed by manual multi-pipetting using an automatic cell counter (Logos Biosystems LUNA-II$^{\rm TM}$) or by flux cytometry (BD FACSAriaIII$^{\rm TM}$) in single-cell mode. When prepared for FACS seeding, cells were rinsed twice with PBS as usual, then resuspended in N2B27 at a concentration of $5\mathrm{-}10.10^6$cells/mL, and strained using cell strainers (Falcon® 352235). Exposure to the GSK3 inhibitor CHIR 99021 (a Wnt agonist called \textit{Chi} throughout the text) during 48 and 72 hrs starts the elongation process. The error in initial seeding number for both methods (Fig.~S1A) was measured by manually counting the number of cells in each individual well between $30\,\rm{min}$ up to $1\,\rm{h}$ after seeding. This way, cells had sedimented but had not started to aggregate.

\subsection{Immunofluorescence staining}

Gastruloids were collected in a 15$\,$mL Falcon tube precoated with PBSF (10\%~FBS in PBS with $\rm{Mg^{2+}}$ and $\rm{Ca^{2+}}$) with a cut P1000 tip and washed once with PBS. Gastruloids were fixed in 4\% PFA for $2\,\rm{h}$, washed once with PBSF and twice with PBS, then resuspended in 1$\,$mL PBS. After this step, they are stored at 4°C (up to several months). For immunofluorescence staining procedure, gastruloids were permeabilized in 10$\,$mL of PBSFT (10\% FBS + 0.03\% Triton in PBS with $\rm{Mg^{2+}}$ and $\rm{Ca^{2+}}$) for $1\,\rm{h}$ at RT; incubated overnight at 4°C in 0.5$\,$mL PBSFT with 1$\,$µL of 1$\,$mg/mL DAPI and with the diluted primary antibody added (see Table~S2 for primary antibody information). Gastruloids were washed 3 times with 10$\,$mL of PBSFT, each wash for $30\,\rm{min}$ at RT. They were incubated overnight at 4°C in 0.5$\,$mL PBSFT with 1$\,$µL DAPI and the secondary antibody diluted 1:500 for anti-rat IgG (Invitrogen A21208) and anti-rabbit IgG (Invitrogen A3157). They were subsequently washed at RT twice in PBSFT each time for $30\,\rm{min}$ and once in PBS for $30\,\rm{min}$ before mounting. All incubations and washes were done on a shaker and all centrifugations were at 10$\,$g for $2\,\rm{min}$ to pellet gastruloids at the bottom of the 15$\,$mL tube. For mounting, gastruloids were transferred for cleaning with a cut P1000 tip in a 6-well dish with 3$\,$mL PBS then transferred to a 1.5$\,$mL low-binding tube to remove all impurities. After removing all PBS, 150$\,$µL of mounting medium (50\% Aqua-Poly/Mount (Polysciences 18606-20) / 50\% PBS with $\rm{Mg^{2+}}$ and $\rm{Ca^{2+}}$) were added and gastruloids were transferred with a cut P200 tip to a glass bottom dish (Cellvis D35-10-1.5-N). Finally, after rearranging the gastruloids to avoid any contact, they were covered by a cover glass.

\subsection{Phalloidin staining}

For phalloidin staining, gastruloids were collected, fixed, and stored similarly to the immunofluorescence staining protocol, except the fixation step was for $1\,\rm{h}$. Fixed gastruloids stored in PBS were transferred to a staining solution containing $1\,$mL of PBST (0.1\% Triton in PBS with $\rm{Mg^{2+}}$ and $\rm{Ca^{2+}}$) with 1$\,$µL of 1$\,$mg/mL DAPI and $2.5\,$µL 400x phalloidin. The tube was placed at 4°C on a shaker overnight. Gastruloids were subsequently washed three times with $10\,$mL of PBST. During each wash, the tube was placed under agitation at 4°C for $20\,\rm{min}$, centrifuged for $2\,\rm{min}$ at $10\,$g, and the supernatant was aspirated. Finally, the samples were mounted on microscopy coverslips with SlowFade$^{\rm TM}$ Glass Antifade mounting medium (Invitrogen S36917) using homemade spacers.

\subsection{Confocal imaging}

Gastruloid fluorescence imaging was performed on an LSM900 Airyscan 2 microscope equipped with an Airyscan detector with GaAsP-PMT (Zeiss). We used Airyscan mode to speed up acquisitions and checked that gene expression profile variability and precision were identical in both Airyscan and confocal modes. Before image analysis, raw acquisitions are subjected to Airyscan processing by Zen 3.3 software (Zeiss). Gastruloids were imaged using a 10$\times$ 0.45~NA air objective (Zeiss) with a zoom setting between 1 and 1.7. $150\,$µm thick z-stacks of 30 slices with a voxel size of $220 \times 220 \times 5,000\, \rm{nm^3}$ were acquired, encompassing the lower half of the gastruloid (set on an inverted microscope objective). $405\,$nm, $488\,$nm, and $639\,$nm laser lines were used to image DAPI, Alexafluor-488 (for SOX2), and Alexafluor-647 (for CDX2/BRA/FOXC1), respectively. Gastruloids with phalloidin staining were imaged using a 40$\times$ 1.43~NA oil immersion objective (Zeiss) with a voxel size of $76 \times 76 \times 190\,\rm{nm^3}$. The size of the z-stack was adjusted for each gastruloid to cover the full width of the specimen. The 405$\,$nm and 561$\,$nm lasers were used to excite the DAPI and phalloidin fluorophores, respectively.

\subsection{Image analysis} 

\subsubsection{Morphological analysis: midline length determination}

The main body axis of each gastruloid (i.e., pseudo-AP axis) was defined by computing the medial axis and extending it with straight lines, tangent to each end of the medial axis (Fig.~S1C). The intersection between this extension and the contour defines the anterior and posterior gastruloid tips. The length of the midline is defined as the length of the curved segment between these two tips. 

\subsubsection{2D volume reconstruction}

An equal number of equidistant points were placed along both sides of the 2D contour (left and right contour sides were defined as segments between anterior and posterior tips). Gastruloids were segmented into bins by pairwise joining equivalent points on the left and right sides. From this midsection plane, an approximate volume was reconstructed assuming an approximation of radial symmetry. The volume of the most extreme bins at the tips was computed using the equation for a sphere cap volume. The volume of all other bins was computed assuming a truncated cone volume, with the following formula ($n_\textbf{B}$ is the total number of bins including caps, $h_i$ the width of bin $i$ along the medial axis and $R_i$ half of its length): 
\begin{equation}
  \begin{aligned}
    V = \frac{1}{3}\pi[h_0^{2}(3R_0 + h_0) + \sum_{i=1}^{n_\textbf{B}-1}h_i(R_i^2 + R_iR_{i+1}) + h_{n_\textbf{B}}^{2}(3R_{n_\textbf{B}} + h_{n_\textbf{B}})]
  \end{aligned}
\end{equation}
This pipeline was applied either to images taken in brightfield at the focal plane for live gastruloids (plane of maximal area) or on 2D maximal projections of the DAPI channel for fixed gastruloids. Using 3D segmentation (see below), which provides a more accurate measurement of the volume but also a more elaborate imaging protocol, we estimated the measurement error of this 2D-based volume reconstruction to be 3--20\% (i.e, an overestimation of the actual gastruloid volume depending on the state of elongation (see Fig. S7B)). All volumes obtained by 2D volume reconstruction were corrected for this systematic error.




\subsubsection{3D cell segmentation}

3D segmentation on image stacks of small gastruloids (initial number of seeded cells smaller or equal to 300) was carried out using Cellpose~\cite{Stringer2021}, a state-of-the-art neural network-based segmentation framework. A Cellpose model was trained and fine-tuned on manually segmented 2D images extracted from z-stacks of fixed and stained gastruloids with dual labels for actin and cell nuclei (Fig.~S7A, Methods~D). After 3D segmentation, cell masks were analyzed to filter out poorly-segmented cells and noisy pixels erroneously identified as cells. To reject noisy masks, a bimodal Gaussian mixture model was fitted to the distribution of the logarithm of single-cell volumes (Fig.~S7E). Only cells belonging to the main mode with the highest mean were kept. The resulting filtered segmentation was checked to be in accordance with the imaging data. Finally, \textit{morphological closing} was performed on the individual cell masks to avoid holes and cell-in-cell detections.

\subsubsection{1D Gene expression profile extraction}

For fixed gastruloids, a 2D maximum projection was calculated for each channel across a $150\,$µm thick stack. The morphological slicing used for 2D volume reconstruction was performed on the DAPI channel with $n_B = 200$~bins (for $\overline{N}_0=300$, at $120\,\rm{h}$, each bin is $\sim$$3\,$µm wide along the gastruloid midline). The maximum projected immunofluorescence intensity was averaged over each bin (averaged over bin size, i.e., $\sim$5500 pixels) for each channel to obtain the raw profiles of intensity as a function of the fractional position $x/L$. Any other analysis on the profiles was performed for $0.1\le x/L \le 0.9$ (Fig.~S3H--K). 

As an alternative, for Fig.~S3H only, the slicing was done using three points: equidistant along both sides of the contour and additionally equidistant points along the midline. Each bin limit was computed from a triplet of corresponding points using a second-order polynomial fit (after a coordinate rotation).

\subsection{Growth analysis} 

To assess the reproducibility and scaling of gastruloid growth, midline length and volume were measured every $24\,$h. Gatruloids are either followed individually (Fig. 1A and S2A) or by averaging the midline length and volume of multiple gastruloids (Fig.~1B). The total cell count $N$ is calculated using the proportionality between $V$ and $N$ (Fig.~1C and S7) or obtained by direct measurement via chemical dissociation (Fig.~1B inset, Methods K). Exponential growth is assumed for the cell count $N$ for individual (Fig.~S2B) and average growth curves. It can be expressed as follows with $N_0$ the initial number of seeded cells and $t_{\rm{D}}$ the effective doubling time :
\begin{equation}
  \begin{aligned}
   N(t)=N_0e^{\frac{t\rm{ln}(2)}{t_{\rm{D}}}}
  \end{aligned}
\end{equation}
The effective doubling time $t_{D}$ was extracted from growth curves via linear fitting of $\rm{ln}(N(t))$ (Fig.~S2B and S2E). We used error propagation to compute the expected error on the cell number $\Delta N$ due to the fluctuations in the initial number of seeded cells $\Delta N_0$ and effective doubling time $\Delta t_{\rm{D}}$:
\begin{equation}
  \begin{aligned}
   \left( \frac{\Delta N}{N} \right)^2 = \left( \frac{\Delta N_0}{N_0} \right)^2 + \left( \frac{t\rm{ln}(2)\Delta t_{\rm{D}}}{t_{\rm{D}}^2} \right)^2
  \end{aligned}
\end{equation}

\subsection{Positional error analysis}

\subsubsection{Determination of boundary position}

The maximum and minimum intensity of each profile, respectively $I_\mathrm{max}$ and $I_\mathrm{min}$, were determined by calculating the average intensity in the 10\% of bins with respectively the highest and the lowest values (Fig. S5A) \cite{Dubuis2013}. The profile is spline-fitted for $0.1\le x/L \le 0.9$ and the position at which I is equal to $(I_\mathrm{max}+I_\mathrm{min}) /2$ defines the individual profile boundary position $x/L_\mathrm{EC50}$ (Fig. S5A). For better automation, the boundary position is searched in a gene-specific region, e.g. with $0.4\le x/L \le 0.9$ for SOX2. 

\subsubsection{Determination of $\sigma_{x/L}$}

For each gene, a generalized version of the positional error is calculated using error propagation~\cite{Dubuis2013a}:
\begin{equation}
\begin{aligned}
\sigma_{x/L} = \sigma_\mathrm{I} \left|\frac{ \mathrm{d\overline{I}}}{\mathrm{d(x/L)}} \right| ^{-1}
\end{aligned}
\end{equation}
The error $\sigma_\mathrm{I}$ is measured as the standard deviation of the raw intensity profiles. The derivative is calculated at each position $x/L$ as the slope of a fitted order-three polynomial on 15 consecutive data points, i.e., bins ($\sim$45$\,$µm or $\sim$3 cells). 

 This value was converted to cell diameter ($d_c$) units using $\epsilon =  \sigma_\mathrm{x/L} L /  d_c$. The measured length $L$ in fixed gastruloids is impacted by the shrinking of gastruloids due to fixation and mounting media and we can write $\epsilon =  \sigma_\mathrm{x/L} L_\mathrm{fixed} / d_c SF$, where $SF$ is the shrinkage factor (Fig.~S6, Methods J). Both the measurements of $d_c$ and $\mathrm{SF}$ come with an error (see below) and by error propagation, we obtain an error on $\epsilon$ that is represented as the width of a gray band in Figures~3C,~S5, and~S8D:

\begin{equation}
    \begin{aligned}
        \Delta \epsilon^2 = 
        ( \frac{1-\mathrm{SF}}{L_\mathrm{fixed}} \Delta d_c)^2 + ( \frac{d_c}{L_\mathrm{fixed}} \Delta \mathrm{SF})^2   
    \end{aligned}
\end{equation}

\subsection{Variance-based analysis of profiles}

\subsubsection{$\chi^2$-minimization }

We minimize the total deviation $\chi^2$ of the intensity profiles from the mean across $n$ gastruloids. We assume that for each profile $I^{(\mu)}$ $(\mu = 1, ..., n)$ we can quantify this deviation for the (unknown) true intensity profile $i^{(\mu)}$  by an additive constant $\alpha_{\mu}$ and a scale factor $\beta_{\mu}$, such that $I^{(\mu)} = \alpha_{\mu} + \beta_{\mu}i^{(\mu)}$. The total deviation $\chi^2$ can be written:

\begin{equation}
    \begin{aligned}
        \chi^2\left( \left\{ \alpha_{\mu}, \beta_{\mu} \right\} \right) = \sum_{\mu=1}^{n}\int_{0}^{1}\rm{d}x\left( \left[ \frac{I^{(\mu)} - \alpha_{\mu}}{\beta_{\mu}} \right]- \overline{i}\right)^2
    \end{aligned}
\end{equation}
We minimized $\chi^2$ to learn $\alpha_{i}$ and $\beta_{i}$ for each intensity profile (either individual in Fig. S5C or mean in Fig. 4B, S8E and F) under the constraints $\sum_{\mu=1}^{n}\alpha_{\mu}=0$ and $\prod_{\mu=1}^{n}\beta_{\mu}=1$.

\subsubsection{Variance decomposition}

In the case of the comparison of mean profiles for different plates, we used variance decomposition (Fig.~S8E) to estimate which part of the variance was due to intrinsic versus inter-plate variance: 

\begin{equation}
    \begin{aligned}
        \sigma_\mathrm{tot}^2 
        & =  \frac{1}{\sum{\mathrm{n_i}}} \sum_\mathrm{i=1}^{n_p} n_i ( \overline{I}_i - <\overline{I}>)^2
        + \frac{1}{\sum{\mathrm{n_i}}} \sum_\mathrm{i=1}^{n_p} \mathrm{n_i} \sigma_{I_i}^2
    \end{aligned}    
\end{equation}

where $n_p$ is the number of plates and $n_i$ the number of gastruloids in plate $i$. The first term represents the inter-plate (variance of the means) and the second term represents the intra-plate variance (mean of variances). 

The results of this decomposition depend on the position $x/L$ along the midline of the gastruloid. To extract a single inter-plate component of the variance per gene and experiment (Fig.~S8F), we calculated a weighted average of $\sigma_\mathrm{inter}$: 

\begin{equation}
    \begin{aligned}
        \overline{\sigma}^2_\mathrm{inter} 
        = \frac{\sum_x \sigma^2_\mathrm{inter}(x) \mathrm{I}(x)}{\sum_x I(x)} 
    \end{aligned}    
\end{equation}


\subsection{Determination of the shrinkage factor}

The various steps of the immunofluorescence and phalloidin staining protocols affect the geometry of the gastruloid isotropically. In particular, the fixation and the mounting media tend to shrink gastruloids. We determined the shrinkage factor by comparing the 2D reconstructed volume of mounted gastruloids from images obtained by confocal microscopy to the volume of the same gastruloids imaged in brightfield just before collection. A one-dimensional shrinkage factor was defined by the ratio : $SF = 1-(V_{IF}/V_{BF})^{1/3}$. This factor quantifies by how much gastruloid size is reduced during the staining protocol. As the staining protocols are done in batches, we cannot calculate a shrinkage factor per individual gastruloid. We calculated $SF$ on the average volume before and after protocol for a given plate (Fig.~S6A). For $50\%$ PBS and $50\%$ aqueous mounting medium (Aqua-Poly/Mount, Polysciences 18606-20) used in the immunofluorescence staining protocol, the shrinkage factor is $SF = 0.35 \pm 0.03$ (Fig.~S6B). For SlowFade$^{\rm TM}$ Glass Antifade mounting medium (Invitrogen S36917) used in the phalloidin staining protocol, the shrinkage factor is $SF = 0.36 \pm 0.1$ (Fig.~S6C). In addition, for one dataset, we monitored the evolution of the estimation of the shrinkage factor from 3 days to 3 weeks after mounting and found that it remained constant within this window of time (Fig.~S6D). The shrinkage factor determined for both these mounting media was applied to all measured lengths and volumes from stained gastruloids.

\subsection{Gastruloid cell counting by chemical dissociation}

Individual gastruloid chemical dissociation was carried out on gastruloids seeded using FACS. First, 140$\,$µL of media per well was removed from the 96-well plates where the gastruloids were grown and gastruloids were washed with 200$\,$µL of PBS with $\rm{Mg^{2+}}$ and $\rm{Ca^{2+}}$. Brightfield images of the gastruloids were taken with an Olympus IX83 microscope with a 10$\times$ objective to later determine individual gastruloid volumes via 2D~volume reconstruction (Fig.~S1C, Methods F2). Gastruloids were then washed with 200$\,$µL of PBS without $\rm{Mg^{2+}}$ and $\rm{Ca^{2+}}$ and transferred to a 96~flat bottom well plate (TPP$^{\rm TM}$ 92096) with a cut P100 tip coated with FBS to prevent gastruloid deformation and sticking. 160$\,$µL of media was removed and 80$\,$µL of 10X Trypsin was added per well. The plate was incubated at 37°C for 5 min. Next, the solution in each well was pipetted up and down to dissociate each gastruloid into a single-cell suspension. The plate was again incubated at 37°C for 5$\,$min. 130$\,$µL of FBS was added per well to neutralize the Trypsin. To continue dissociating the gastruloids, the solution in each well was again pipetted up and down with the same pipette tips to avoid cell loss. The cells in each well were then fixed in 1\% PFA for 15 min and stained with a solution of 0.5\% Triton and 0.2\% DAPI. The wells were imaged using the Zeiss LSM900 with a 10$\times$ 0.45~NA air objective with z-stacks of $60\,$µm and a voxel size of $829 \times 829 \times 7,500\ \rm{nm^3}$. Finally, the cells were automatically detected by watershed segmentation. Individual gastruloid dissociation results agreed with mean cell count data obtained by pooling 48 gastruloids and carrying out bulk dissociation.

\subsection{Equivalent cell diameter determination}

We developed two independent methods to determine gastruloid volume and cell count. The first method involves chemical dissociation based cell counts of individual gastruloids (see Methods~K) and a 2D volume reconstrution from images taken of the gastruloids prior to dissociation.  We applied this protocol to gastruloids with a range of initial number of seeded cells ($\overline{N}_0$ = 100, 300, 500 and 800) and at five daily time points (ranging from $24\,\rm{h}$ to $120\,\rm{h}$). For each gastruloid, we first reconstructed their volume $V$ from a 2D cross-section image at the center of the gastruloid using brightfield microscopy (see Methods F.2.). Subsequently the imaged gastrluoid was dissociated chemically to obtain its cell count $N$ (Fig.~1C and S7C). Gastruloid length, volume, and cell count at $120\,\rm{h}$ are reported in Table~S3. From $V$ and $N$ for each individually dissociated gastruloid, we computed their effective cell volume $V_{c} = \frac{V}{N}$ and effective cell diameter $d_{c} = \sqrt[3]{ \frac{6 V_{c}}{\pi}}$. The effective cell diameter corresponds to the mean of the distribution of $d_{c}$, its error to the standard deviation. Before Chi-pulse, $d_c=16.0\pm0.6\,$µm $(4.0\%, n=206)$; after Chi-pulse, $d_c=13.9\pm0.5\,$µm $(3.8\%, n=286)$ (Fig.~S7D). This is evidence of a Chi-pulse-induced reduction in gastruloids' effective cell size by $\sim$13\% (linear dimension). 

The second method is based on 3D cell segmentation (Methods~D and~F.3.). We used single-cell data to extract the volume (sum of single-cell volumes) and cell count (number of 3D segmented cells) from gastruloids aged at least $72\,\rm{h}$ with $\overline{N}_0$ ranging from 50 to 300 cells. A comparison of $V$ and $N$ from these two methods is presented in Fig.~S1F (with extensive additional information in the caption). With the 3D segmentation method, $d_c=13.1\pm0.5\,$µm $(4.0\%, n=108)$. Taking into account the different sources of error and our two independent methods of determination of the effective cell diameter, the relevant linear size of the system at $120\,{\rm h}$ is $d_c=13.5\pm0.8\,$µm.

Note, in our 3D segmentation analyses, we observe small amounts of extracellular space that accumulates during gastruloid formation. These lumen-like structures occur in some gastruloids, but their size is negligible compared to the overall size of the gastruloid. We estimate the amount of extracellular space to be less that one percent of the total gastruloid volume.


\newpage

\section{Supplemental Figures}

\vspace{1cm}

\renewcommand\thefigure{S\arabic{figure}}  
\renewcommand\theequation{S\arabic{equation}}
\setcounter{figure}{0}


\begin{figure}[h!]
    \centering
    \includegraphics[width=0.9\linewidth]{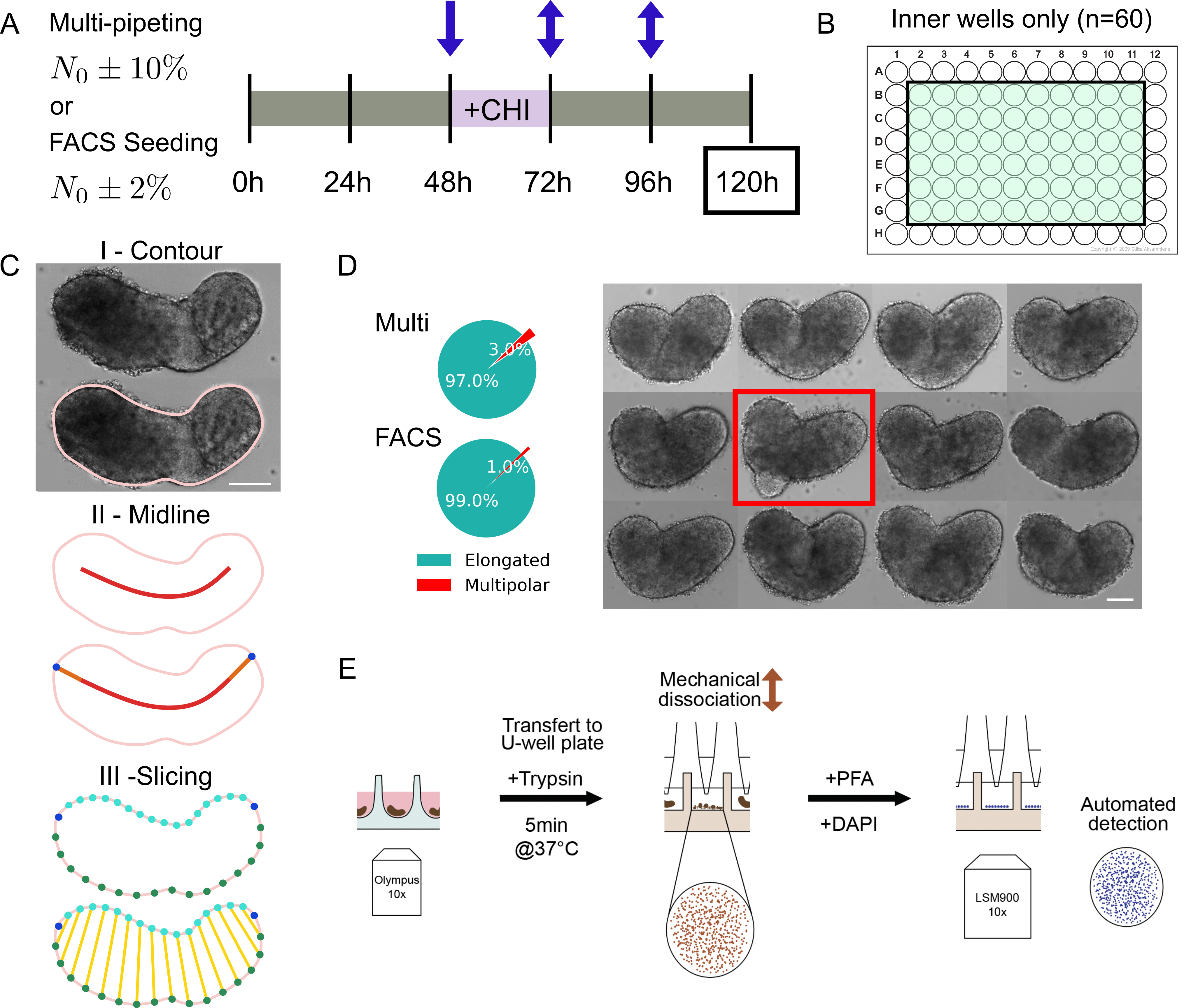}
    \caption[Experimental detail, protocols, and analysis]{Experimental detail, protocols, and analysis. 
    {\bf A}:~Gastruloid protocol as described before with a Chi-pulse on day three~\cite{Beccari2018a}. Initial seeding either done by manual multi-pipetting or using Fluorescence-activated Cell Sorting (FACS)~\cite{VandenBrink2020}, implying a different variability in the initial number of seeded cells $N_0$; 10\% vs. 2\%, respectively. Blue arrows indicate addition of Chiron and change of medium.
    {\bf B}:~Discarding all gastruloids grown in outer wells for increasing reproducibility. Empirical observation determined largely from different behaviors for gastruloids grown in inner versus outer wells~\cite{Mansoury2021}.
    {\bf C}:~Image analysis steps include the definition of a smooth contour (I), drawing the midline (II), and slicing along this midline using an equidistant positioning of two sets of equal-number points on each side of the contour (III). For III, the points in left half (light blue) and in right half (dark blue) are equidistant along the contour, respectively. Gastruloid volume is reconstructed by assuming each slice is rotationally symmetric (i.e., a truncated cone). Scalebar is $100\,$µm.
    {\bf D}:~Gastruloids imaged with brightfield microscopy. Gastruloid elongation efficiency is 97\% for multi-pipetting and 99\% for FACS seeding, for $\overline{N}_0$. The remaining gastruloids have multiple poles (e.g., red framed image). Scalebar is  $100\,$µm.
    {\bf E}:~Schematic of the protocol to measure the volume and cell count of individual gastruloids. Brightfield images of gastruloids are acquired before chemical dissociation, left; fluorescent images of \textit{all} individual cells composing the gastruloid are acquired after dissociation using confocal microscopy (see Methods).
    }
    \label{figs:gastruloidprotocol}
\end{figure}


\newpage
\begin{figure}
    \centering
    \includegraphics[width=0.96\linewidth]{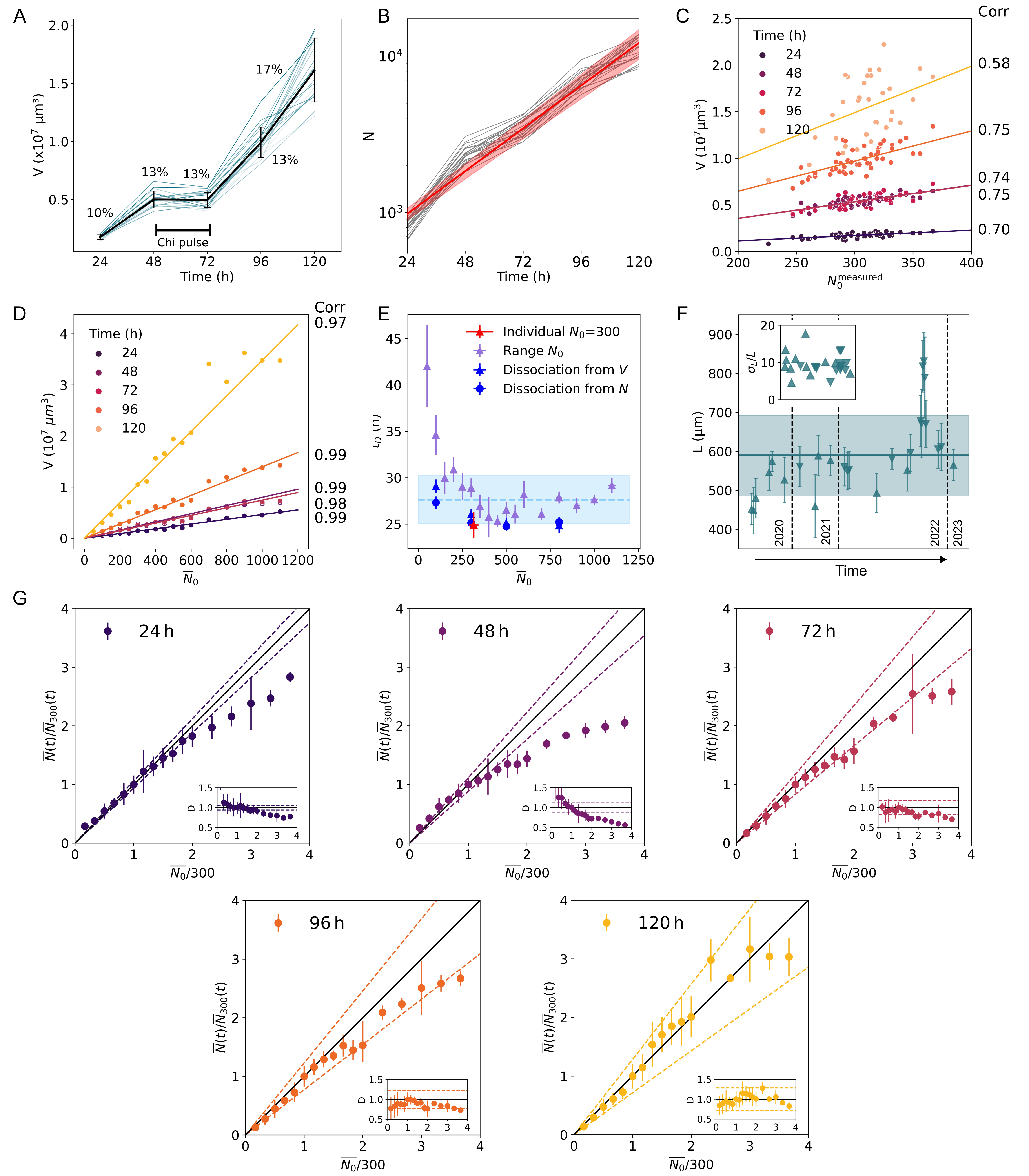}
    \caption[Growth reproducibility and scaling]{Growth reproducibility and scaling. 
    [Caption see next page.]}
\end{figure}

\clearpage
\begin{figure}
    \centering
    \begin{justify}
    FIG. S2. Growth reproducibility and scaling.
    {\bf A}:~Gastruloid volume as a function of time. Volumes are obtained from 2D reconstruction in Fig.~S1C. Curves shown for 23 gastruloids (subset of Fig.~1A) followed over time individually (blue) and mean (black). Percent variation around the mean is reported for each time point. 
    {\bf B}:~Exponential growth of the total number of cells in individual gastruloids (same as in A). The total cell count $N$ shown in log-scale as a function of time $t$ is obtained from the proportionality between $V$ and $N$ (Fig.~1C and S7). Exponential growth (see Methods) is assumed for each individual growth curve (in grey) to extract the effective doubling time $t_{D}$ for each gastruloid (via linear fitting). Red line corresponds to exponential growth with mean effective doubling time $t_{\rm D}=26.4\pm1.7\,$h. Red shaded area was computed from error propagation.
    {\bf C}:~Gastruloid volumes correlate with $N_0$ at all time points. Scatter plot of individual gastruloid volumes from A at different time points versus $N_0$, measured just after seeding, overlaid by a linear regression fit. The correlation coefficient for each fit is reported on the right y-axis. 
    {\bf D}:~Scatter plot of mean gastruloid volume at different time points versus $\overline{N}_0$, measured just after seeding, overlaid by a linear regression fit. These are the same gastruloids shown in Fig.~1B. Right y-axis shows the correlation between volume and $\overline{N}_0$ for different time points (color). When scanning a large range of average $\overline{N}_0$ ($50\le \overline{N}_0 \le 1100$), the correlations increase significantly.
    {\bf E}:~Effective doubling time $t_{\rm D}$ as a function of $\overline{N}_0$. The effective doubling time is obtained by fitting  growth curves of the number of cells by an exponential growth model (see Methods). For round markers, $t_{\rm D}$ is extracted from cell counts measured directly by chemical dissociation. For triangle markers, cell counts are obtained from volume measurements using the relationship in Fig.~1C. Red markers correspond to the individual gastruloids in Fig.~1A; purple markers correspond to averaged data in Fig.~1B; blue markers to the inset in Fig.~1B. Average effective doubling time for gastruloids seeded with $150\le \overline{N}_0 \le 1100$ is $\overline{t}_{\rm D}=27.6\pm2.6\,{\rm h}$ (mean as blue dashed line; light blue area standard deviation).
    {\bf F}:~Evolution of average midline length per experiment over three years (2020--2023) for gastruloids with $\overline{N}_0=300$ at $120\,\rm{h}$. Downward triangles are experiments seeded by multi-pipetting; upward triangles are experiments seeded using FACS. Each experiment had between 30--40 samples. Error bars are standard deviations across individual samples. The blue line represents the overall average across all experiments with blue shaded area as the standard deviation: $\overline{L}=590 \pm 102\,$µm $(17\%,\ n=30)$. Inset shows the corresponding evolution of the variability of the mean gastruloid midline length per experiment. Intra-experiment variability in length is on average $\langle \sigma_{L}/L\rangle = 9.4 \pm 2.7 \%\ (n=30)$. Over three years, both the gastruloid midline length and its variability are highly consistent.
    {\bf G}:~Average cell count $\overline{N}(t)/\overline{N}_{300}(t)$ as a function of the initial average seed cell count $\overline{N}_0/300$ in units of the average reference seed cell count $\overline{N}_0=300$. Five panels correspond to gastruloid ages at 1 through 5 days (also encoded by color). Black diagonal (slope = 1) represents perfect scaling (see main text) of gastruloid size at time $t$ upon changes in $\overline{N}_0$ ranging over $50\le \overline{N}_0 \le 1100$. For each time point, using a simple exponential growth model, the dashed lines estimate the bounds on the expected deviations from perfect scaling due to fluctuations in both $\overline{N}_0/300$ and in the doubling time $t_{\rm D}$ (E and Methods). Insets show deviations $D$ from perfect scaling: $ D= \frac{\overline{N}(t)/\overline{N}_{300}(t)}{\overline{N}_0/300}$, as a function of the initial average seed cell count $\overline{N}_0/300$ in units of the average reference seed cell count $\overline{N}_0=300$. Black horizontal line represents perfect scaling and the dashed lines show expected deviations from error propagation.     
    \end{justify}
    \label{figs:gastruloidcontrol}
\end{figure}


\newpage
\begin{figure}
    \centering
    \includegraphics[width=\linewidth]{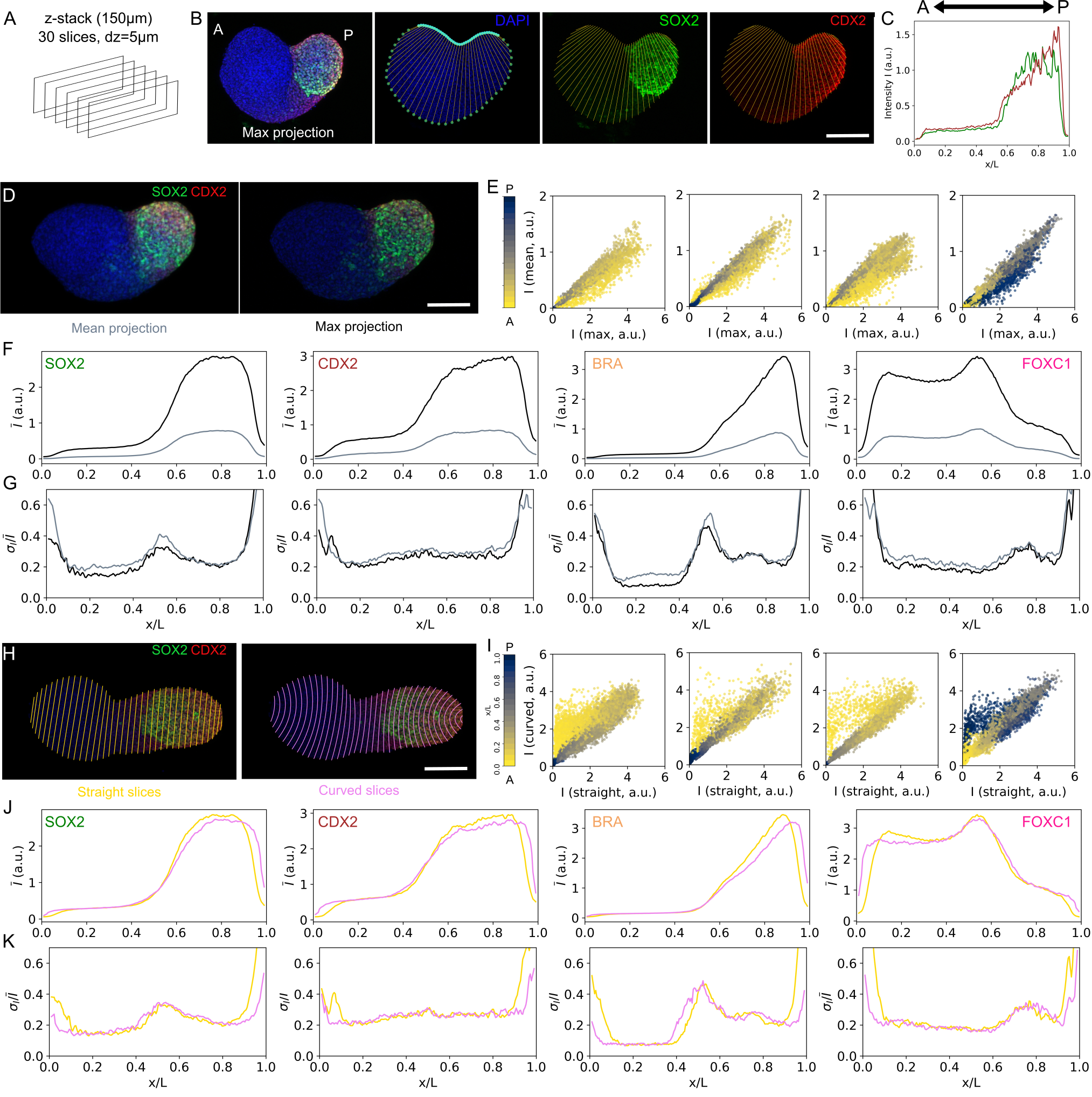}
    \caption[Immunofluorescence image analysis]{
    Immunofluorescence image analysis. [Caption see next page.]}
\end{figure}

\clearpage
\begin{figure}
    \centering
    \begin{justify}
    FIG. S3. Immunofluorescence image analysis.
    {\bf A}:~Fixed gastruloids are imaged by confocal microscopy in z-stacks of 
    $150\,$µm (30 slices, $\rm dz=5\,$µm).   
    {\bf B}:~Analysis pipeline of Fig.~S1C is applied to the DAPI channel for each gastruloid to extract midline, contour, and equidistant slices. Fluorescence intensities of the other channels are max projected (here illustrated with SOX2 (green) and CDX2 (red)) and intensities of individual slices are integrated to obtain a single value per slice and to construct one-dimensional expression profiles as a function of slice position along the midline.  Scalebar is $100\,$µm.
    {\bf C}:~One-dimensional profiles of SOX2 (green) and CDX2 (red) along the midline obtained for the gastruloid in B. 
    {\bf D}:~Visual comparison of mean (left) versus maximum (right) projection of a gastruloid stained for SOX2 (green) and CDX2 (red).  Scalebar is $100\,$µm.
    {\bf E}:~Quantitative comparison of maximum (x-axis) versus mean (y-axis) projection of intensities for the four examined genes in individual gastruloids from Fig.~2 (n=\{44, 44, 48, 46\} respectively for SOX2, CDX2, BRA and FOXC1). Color code corresponds to the position of each slice along the midline (yellow towards the anterior pole, gray-blue towards the posterior pole). 
    {\bf F}:~Mean profiles of expression of the four genes as a function of  relative position $x/L$ using either maximum (black) or mean (gray) projection. 
    {\bf G}:~Variability as a function of the relative position $x/L$ along the midline of each set of gastruloids for the four genes. Gray and black lines correspond to the variability computed respectively from either mean or maximum projections. Measured variability is lower when using maximum projection. 
    {\bf H}:~Visual comparison of gastruloid slicing methods, straight lines (left, yellow) versus curved lines (right, pink); immunostained gastruloid stained for SOX2 (green) and CDX2 (red). Straight lines are line segments calculated between the equidistant points along both sides of the contour as in Fig.~S1C. Curved lines are obtained using both equidistant points along the contour and along the midline. From this combination of points, a parabolic equation is calculated using a second-order polynomial fit. This procedure is meant to recapitulate the overall curvature of the gastruloid. 
    {\bf I}:~Quantitative comparison of intensities using straight (x-axis) versus curved (y-axis) line slicing for the four examined genes in individual gastruloids from Fig.~2 (n=\{44, 44, 48, 46\} for SOX2, CDX2, BRA and FOXC1, respectively). Color code corresponds to the position of each slice along the midline (yellow towards the anterior pole, gray-blue towards the posterior pole).
    {\bf J}:~Mean profiles of the four stained sets of gastruloids from Fig.~2 as a function of relative position $x/L$ using either straight (yellow) or curved (pink) line slicing. 
    {\bf K}:~Variability as a function of the relative position $x/L$ along the midline of each set of gastruloids for the four genes. Yellow and purple lines correspond to straight and curved line slicing, respectively. Using the curved lines method diminishes border effects on profiles of the four genes (mean and variability). No significant change is observed for the most part of the gastruloid midline, making both methods essentially equivalent. For computational simplicity, we employ the straight lines method. All profiles are represented for $0.1\le x/L \le 0.9$ in the rest of the paper. 
    \end{justify}
\end{figure}


\clearpage
\begin{figure}
    \centering
    \includegraphics[width=0.9\linewidth]{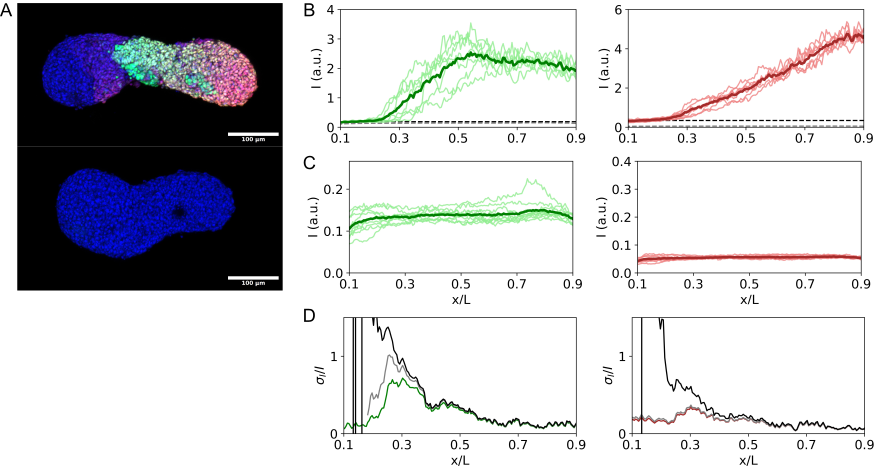}
    \caption[Immunofluorescence background estimation]
    {Immunofluorescence background estimation.
    {\bf A}:~Gastruloids dual-labeled immunofluorescently for SOX2 and CDX2 expression using the regular protocol, as in Fig.~2; bottom gastruloid is missing the primary antibodies to estimate the background noise due to non-specific interactions of the secondary antibodies which are estimated the dominant source of background noise in the staining and imaging procedures~\cite{Dubuis2013}. 
    {\bf B}:~Individual (n=10, light color) and mean profiles (bold) for SOX2 (left, green) and CDX2 (right, red) labeled including primary antibodies. Gray dashed line is the background estimation from C; black dashed line is the background calculated from the raw profiles as the mean intensity level in the $10\%$ region of lowest expression ($I_{min}$). These two dashed lines are confounded in the case of the SOX2 profiles, confirming that the control experiment is a good estimate of .
    {\bf C}:~Control experiment without primary antibodies; individual (n=10, light color) and mean profiles (bold) for SOX2 (left, green) and CDX2 (right, red).
    {\bf D}:~Comparison of the variability ($\sigma_I/\overline{I}$) using either the raw mean profile (bold color), the control-corrected profile (bold grey) or the $I_{min}$-corrected profile (bold black).
    }

\end{figure}


\clearpage
\begin{figure}
    \centering
    \includegraphics[width=0.9\linewidth]{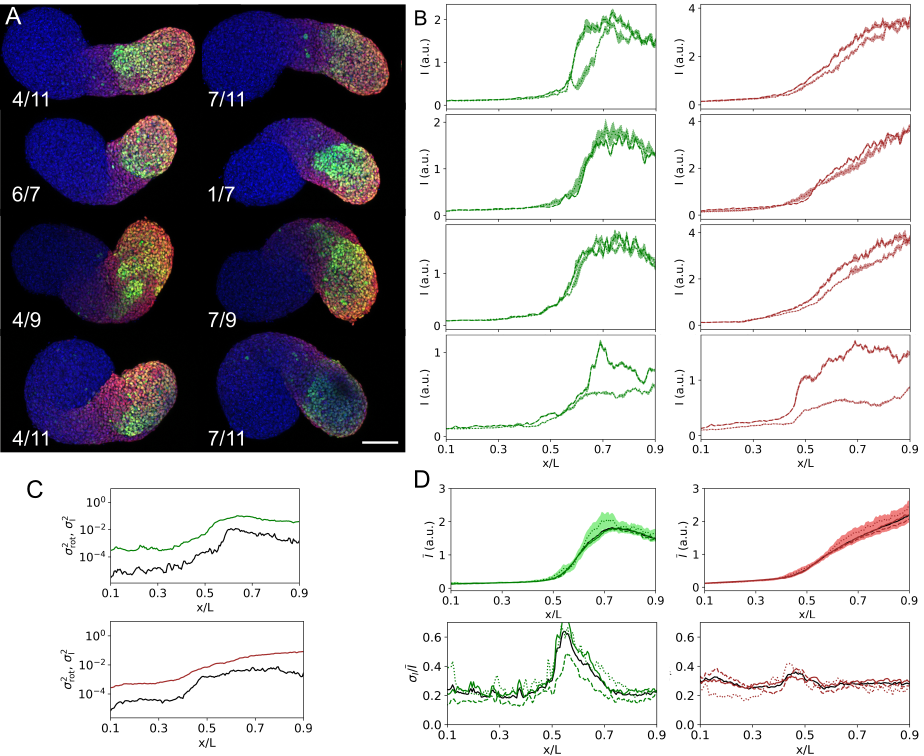}
    \caption[Immunofluorescence measurement error estimation: specimen rotation]
    {Immunofluorescence measurement error estimation: specimen rotation. 
    {\bf A}:~4 single gastruloids immunofluorescently stained for SOX2 and CDX2. Gastruloids are mounted in PBS medium and rotated manually via flushing for each exposure n=7--11 times, and taken from different view angles. Images are categorized for two different orientations of the gastruloid view angle: a "side view" (left column) and a "backside view" (right column). The preferential orientation is determined by the gastruloid shape and is different from gastruloid to gastruloid. Scalebar is $100\,$µm.
    {\bf B}:~Mean profiles of SOX2 (green) and CDX2 (red) expression for the four gastruloids in A gathered by category: side view (dashed line) and backside view (dotted line). Panels in each row correspond to four experiments with a different individual gastruloid (n=\{11, 9, 7, 11\} images, respectively). Shaded areas are standard errors in all graphs. 
    {\bf C}~:~Variance of mean profiles in the four gastruloids due to specimen rotation for SOX2 (black, top) and CDX2 (black, bottom) calculated by bootstrapping the data in B. This variance is compared to the total variance (SOX2: green, top; CDX2: red, bottom) of n=88 gastruloids. The rotation-induced variance represents less than 10\% of the total variance.
    {\bf D}~:~Mean expression profiles (top) and variability (bottom) for SOX2 (green) and CDX2 (red) of the n=88 gastruloids from C classified according to their orientation (i.e., determine AP orientation using SOX2 expression, determine straight versus crescent orientation, determine L/R orientation for crescent shapes; line style as in B). Black lines are the mean profile and standard deviation of gene expression in the total population. This classification based on specimen rotation has minimal effect on the values of mean expression or variability. 
    }
    \label{figs:profileerror}
\end{figure}


\newpage
\begin{figure}
    \centering
    \includegraphics[width=0.98\linewidth]{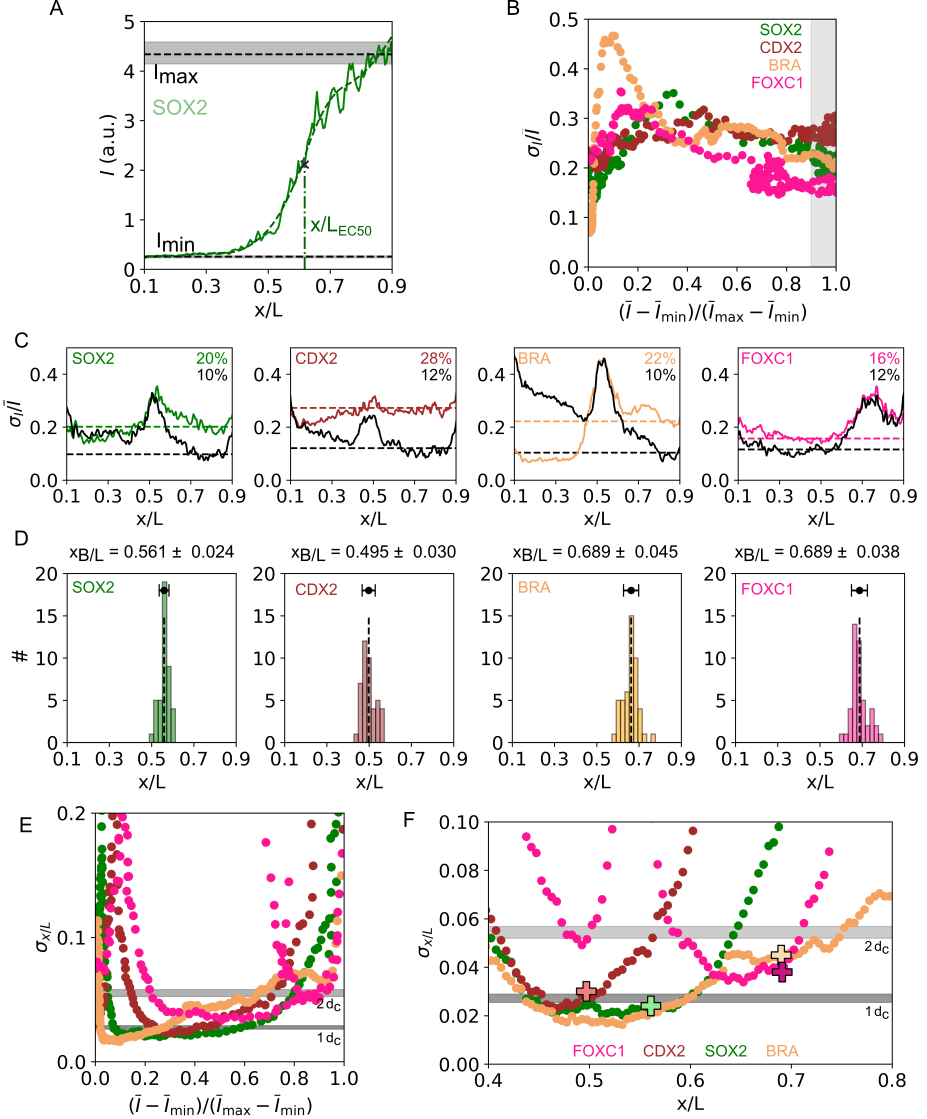}
    \caption[Reproducibility and precision of gene expression profiles]{
    Reproducibility and precision of gene expression profiles. [Caption see next page.]
    }
    \label{figs:statisticalanalysis}
\end{figure}

\newpage
\begin{figure}
    \centering
    \begin{justify}
    FIG. S6. Reproducibility and precision of gene expression profiles.
    {\bf A}:~EC50 determination: $I_{\rm max}$ and $I_{\rm min}$ for individual one-dimensional gene expression profiles are defined as the average value of the $10\%$ largest and lowest expressing bins, respectively. The raw profile (green, plain curve) is spline fitted (green, dotted curve), and the position where the fit is equal to $ (I_{\rm max} + I_{\rm min})/2$ defines $x/L_{\rm EC50}$.
    {\bf B}:~Variability $(\sigma_I/\overline{I})$ as a function of normalized intensity $\overline{I}$. $\overline{I}_{\rm max}$ and $\overline{I}_{\rm min}$ are determined as in A for SOX2, CDX2, BRA and FOXC1. The average value in the gray region (defined by the gene being expressed at more than 90\% of its max level) is used as a measure of gene expression reproducibility for the fully \textit{induced} gene. 
    {\bf C}:~Comparison between raw (colored lines) and $\chi^2$-minimized (black lines) variability as a function of position $x/L$ for SOX2, CDX2, BRA, and FOXC1 for data set in Fig. 2. Dashed lines represent the average variability in the region where genes are most highly expressed (see B). These values decrease from $\sim$20\% to $\sim$10\% after $\chi^2$-minimization, showing the potential for reproducibility after systematic error reduction, similar to what is seen in the fly embryo~\cite{Dubuis2013}.
    {\bf D}:~Distribution of $x/L_{\rm EC50}$ for each of the four markers. The average value is the gene boundary position $x_B/L$ and the standard deviation around this value is a measure of the positional error of the boundary position. 
    {\bf E}:~Generalized positional error as a function of the normalized intensity for each marker (color code as in B). The zones of highest precision (i.e., $\sigma_{x/L}\le 5\%$) correspond to the transition regions between low- and high-expression domains. 
    {\bf F}:~Positional error $\sigma_{x/L}$ calculated for four genes as in Fig.~3C. The positional errors at the boundaries are shown here at the mean boundary position $x_B/L$ extracted in D (big crosses, bootstrapped errors are within marker size). The values from both methods are consistent and for all genes, the positional errors at the boundaries correspond to a linear dimension of 1--2 cell diameters (gray bands).
    \end{justify}
\end{figure}


\clearpage
\begin{figure}
    \centering
    \includegraphics[width=0.8\textwidth]{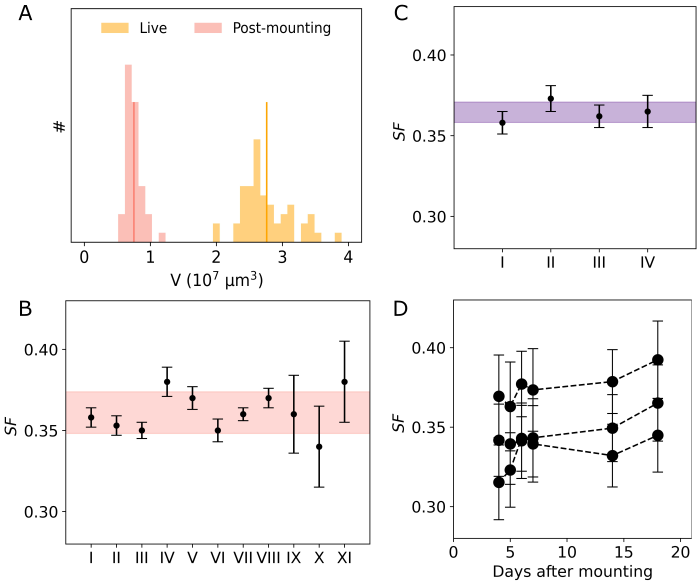}
    \caption[Shrinkage due to fixation and sample mounting]{Shrinkage due to fixation and sample mounting.
    {\bf A}:~Distribution of gastruloid volumes $V_{BF}$ (gastruloids seeded with $\overline{N}_0=300$ cells) at $120\,\rm{h}$; 2D volume reconstruction from either brightfield images or maximum projection of confocal images on the DAPI channel. Gastruloid volumes after fixation and mounting (red, n=47) are $\sim$3 times smaller than the same set of gastruloids imaged live before fixation (yellow, n=52). The number of gastruloids after fixation and mounting is always smaller than during live imaging as gastruloids are lost during the protocol.
    {\bf B}:~A one-dimensional shrinkage factor is defined by the ratio of the average values in A: $SF = 1-(V_{IF}/V_{BF})^{1/3}$. This factor quantifies by how much gastruloid size is reduced during the staining protocol. It is applied to all measured lengths of midlines from stained gastruloids. Gastruloids are mounted in $50\%$ PBS and $50\%$ aqueous mounting medium (Aqua-Poly/Mount, Polysciences). I--XI are 11 independent experiments where $SF$ was calculated on gastruloids initially seeded with $\overline{N}_0=300$ cells and imaged at $120\,\rm{h}$ after seeding. Error bars are from bootstrapping with on average n=51 for live images and n=42 gastruloids after fixation and mounting (experiments I-VIII,) or n=20 for live images and n=10 after fixation and mounting (IX-XI). The shrinkage factor in these experimental conditions is $SF = 0.35 \pm 0.03$.
    {\bf C}:~Same as B for a glycerol-based SlowFade$^{\rm TM}$ Glass Antifade mounting medium (Invitrogen) used in the phalloidin staining protocol. Each data point corresponds to an average gastruloid pool of n=49 for live and n=27 after fixation and mounting. Error bars from bootstrapping. Experiment I corresponds to $\overline{N}_0=100$~cells at $120\,\rm{h}$, experiments II-IV correspond to $\overline{N}_0=300$ at $72\,\rm{h}$, $96\,\rm{h}$ and $120\,\rm{h}$, respectively. The shrinkage factor in this mounting medium is $SF = 0.36 \pm 0.1$. Note that gastruloids are fixed for $1\,\rm{h}$ in the phalloidin staining protocol while they are fixed for $2\,\rm{h}$ in the immunostaining protocol.
    {\bf D}:~Shrinkage factor stability over time for three different mounting techniques in $50\%$ PBS and $50\%$ aqueous mounting medium (Aqua-Poly/Mount, Polysciences): on a slide with a 250$\,$µm spacer or in a glass bottom dish w/ or w/o coverslip. Shrinkage factor measured repeatedly in the same set of gastruloids from IX-XI of B between three days and three weeks. Error bars from bootstrapping.
    }
    \label{figs:shrinkagefactor}
\end{figure}


\newpage
\begin{figure}
    \centering
    \includegraphics[width=\linewidth]{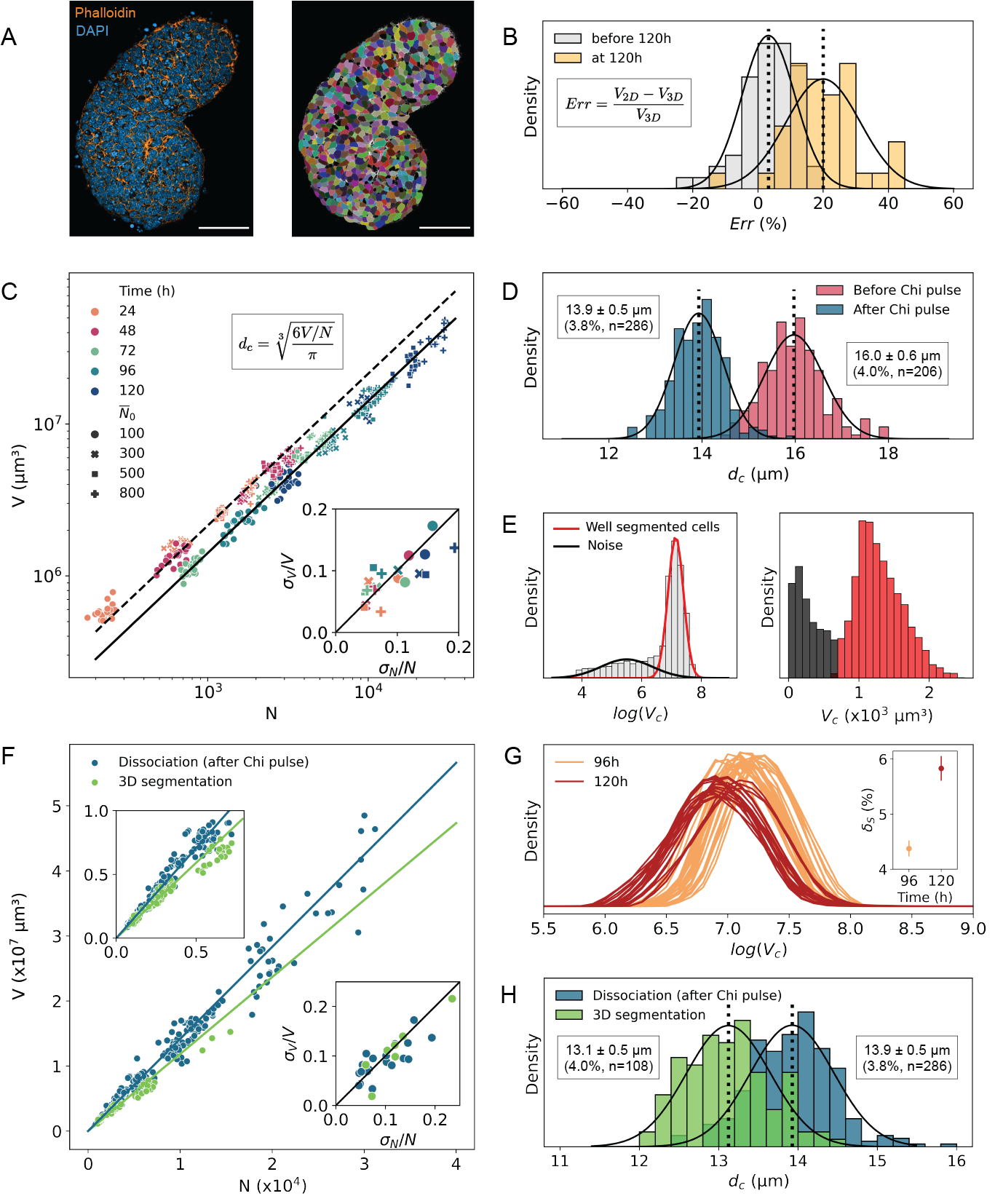}
    \caption[Determination of total cell count and effective cell diameter]
    {Determination of total cell count and effective cell diameter. [Caption see next page.]
    }
    \label{figs:celldiameter}
\end{figure}

\newpage
\begin{figure}
    \centering
    \begin{justify}
    FIG. S8. Determination of total cell count and effective cell diameter.
    {\bf A}:~Visualisation of the cell masks obtained by 3D segmentation~\cite{Stringer2021}. (Left:) Slice of a confocal image z-stack of a $120\,\rm{h}$ old gastruloid, seeded from $\overline{N}_0=100$ cells, stained for phalloidin (orange) and DAPI (blue). (Right:) Phalloidin channel from left in grayscale overlaid with cell masks obtained by 3D segmentation (see Methods). Scalebar is $50\,$µm.
    {\bf B}:~Estimation of the discrepancy between 3D and 2D volume reconstruction. The pipeline presented in Fig.~S1C overestimates gastruloid volumes; we estimate by how much using the volume determined by 3D segmentation as a ground truth. Distribution of the error $Err$ on the volume determined by 2D volume reconstruction $V_{2D}$, before $120\,\rm{h}$ and at $120\,\rm{h}$, overlaid by a Gaussian distribution fit for each distribution. Vertical dashed lines correspond to the mean of each distribution. The ground truth 3D volume $V_{3D}$ was obtained from the 3D segmentation. Before $120\,\rm{h}$, $Err=3.2\pm8.2\%$ $(n=56)$. After $120\,\rm{h}$, $Err=20.0\pm11.2\%$ $(n=40)$. The volume was overestimated in both time classes but more so when the gastruloid elongated. Note that this evaluation of the discrepancy between 3D and 2D volume reconstruction is independent of the shrinkage factor (Fig.~S6) because 3D and 2D volume reconstructions are applied to the same shrunken gastruloid mounted with the phalloidin staining protocol.
    {\bf C}:~Scatter plot of the measured volume from 2D reconstruction $V$ (corrected for the error determined in B) versus the total cell count $N$ obtained by chemical dissociation (with the protocol in Fig.~S1E), for 492 individual gastruloids at different time points (color code) and with varying $\overline{N}_0$ (symbol). From $V$ and $N$ for each individually dissociated gastruloid an effective cell volume $V_{c} = V/N$ was computed, and from there we obtain the slope (black lines). The mean $\overline{V}_{c}$ for gastruloids aged from 24 to $48\,{\rm h}$ (before Chi-pulse) and the mean $\overline{V}_{c}$ for gastruloids aged from 72 to $120\,{\rm h}$ (after Chi-pulse) correspond to dashed and full lines, respectively. Inset shows correlation ($r=0.78$) of variability for $V$ and $N$ for sets of gastruloids with identical age and $\overline{N}_0$. The effective cell diameter $d_c$ can be obtained from the distribution of $V_{c}$, or directly from the slopes (see Methods and D).
    {\bf D}:~Distribution of the effective cell diameters $d_c$ per dissociated gastruloid, calculated from each effective single cell volume (V/N), before (red) and after (blue) Chi-pulse. Black lines are a Gaussian fit for each distribution. Vertical dashed lines correspond to the mean of each distribution. Before Chi-pulse, $d_c=16.0\pm0.6\,$µm $(4.0\%, n=206)$; after Chi-pulse, $d_c=13.9\pm0.5\,$µm  $(3.8\%, n=286)$. This is evidence of a Chi-pulse-induced reduction in gastruloids' effective cell size by $\sim$13\% (linear dimension).
    {\bf E}:~Single cell volume distributions serve to reject noisy masks from 3D segmentation results. After an initial rejection of any 3D masks smaller than $10^{4}$ voxels, a bimodal distribution of the logarithm of single cell volumes $V_{c}$ (obtained by 3D segmentation of a $120\,\rm{h}$ old gastruloid with $\overline{N}_0=100$) is fit by a two-component Gaussian mixture model (left). The mode in black corresponds to the distribution of small noisy masks, the mode in red corresponds to the distribution of well-segmented cells. \textit{Morphological closing} is performed on the latter and the corresponding distribution of single cell volumes $V_{c}$ is shown in right panel, with noisy masks (black) and well-segmented masks (red).
    {\bf F}:~Scatter plot of gastruloid volume versus total cell count obtained by two independent methods. Blue: chemical dissociation and 2D volume reconstruction (for gastruloids dissociated after Chi-pulse only). Green: 3D segmentation for volume and cell count measurement (well-segmented cells only, see E). Slope of blue and green lines correspond to the mean $V_{c}$ for chemically dissociated and 3D segmented gastruloids, respectively. Upper left inset shows a close-up for small $V$ and $N$. Lower right inset shows correlation of variability for $V$ and $N$ for both methods. Note that the main error attached to the 3D segmentation volume is due to the estimation of the shrinkage factor of the mounting medium used in the phalloidin staining protocol (Fig.~S5C). 2D volume reconstruction from dissociated gastruloids is applied to images of live gastruloids (i.e., they are not shrunken).
    {\bf G}:~Distribution of the logarithm of single cell volumes $V_{c}$ obtained by 3D segmentation after filtering and reconstruction for $96\,\rm{h}$ (n=28) and $120\,\rm{h}$ (n=20) old gastruloids with $\overline{N}_0=100$. Inset shows dispersion self-similarity $\delta_{S}$, defined as $\langle \sigma_{\rm{log}(V_C)}/\rm{log}(V_C)\rangle$ for each set of distributions. It demonstrates the reproducibility of the dispersion in cell size in individual gastruloids and a further reduction in gastruloid cell size during the elongation process. The low variability indicates that the dispersion is highly conserved across gastruloids.
    {\bf H}:~Distribution of the effective cell diameter per gastruloid, obtained by chemical dissociation (only data from gastruloids dissociated after Chi-pulse) and 3D segmentation, overlaid by a Gaussian fit for each distribution. Vertical dotted lines correspond to the mean of each distribution. With the dissociation protocol, $d_c=13.9\pm0.5\,$µm $(3.8\%, n=286)$. With the 3D segmentation method, $d_c=13.1\pm0.5\,$µm $(4.0\%, n=108)$. Taking into account the different sources of error and our two independent methods of determination of the effective cell diameter, the relevant linear size of the system at $120\,{\rm h}$ is $d_c=13.5\pm0.8\,$µm.
    
    \end{justify}
\end{figure}


\newpage
\begin{figure}
    \centering
    \includegraphics[width=\linewidth]{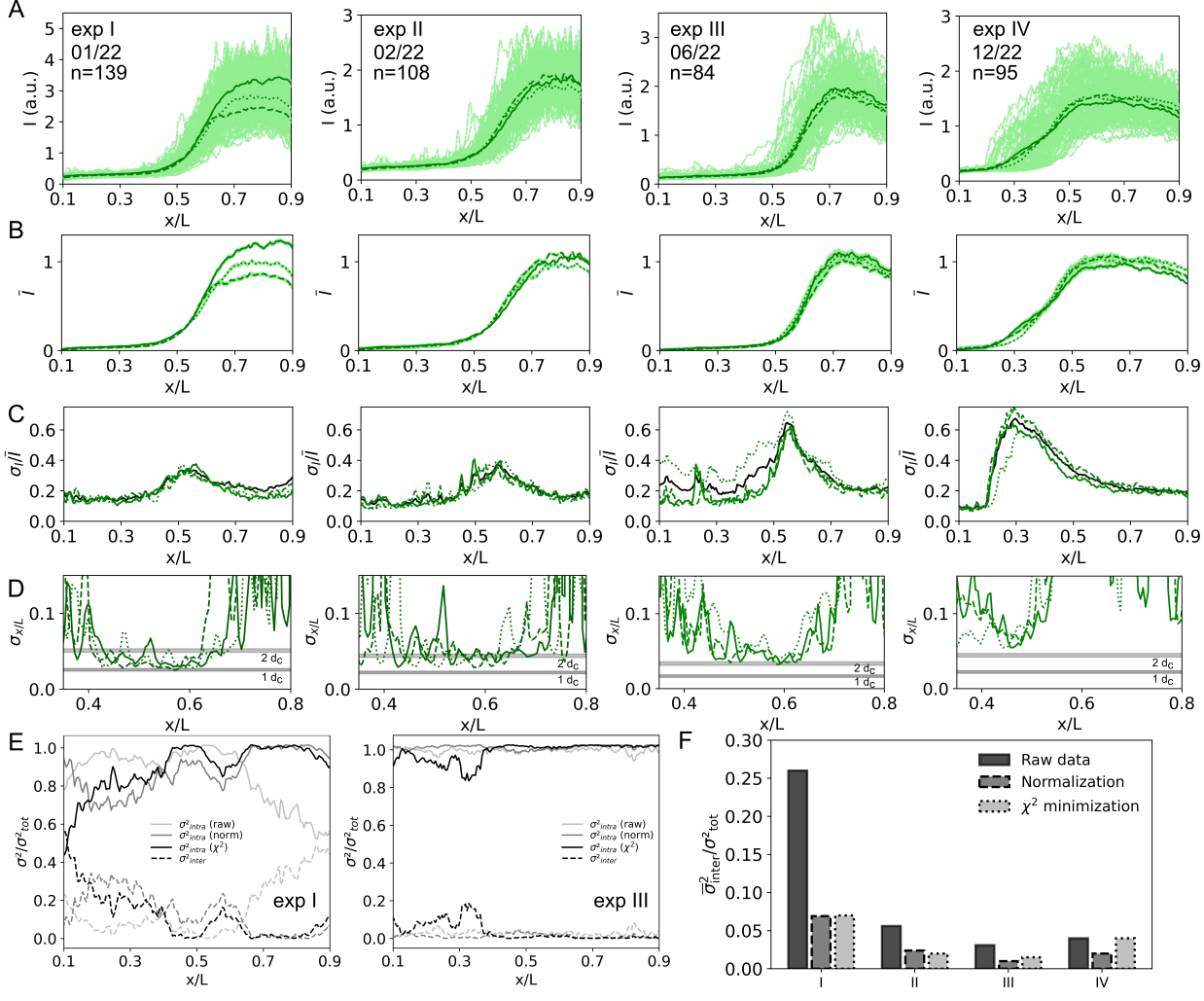}
    \caption[Repeatability and reproducibility of a single experiment]{
    Repeatability and reproducibility of a single experiment. [Caption see next page.]}
\end{figure}

\clearpage
\begin{figure}
    \centering
    \begin{justify}
    FIG. S9. Repeatability and reproducibility of a single experiment. 
    {\bf A}:~Twelve repetitions of the same experiment on different dates (exp I--exp IV, month/year, with n=139, 105, 84 and 95 gastruloids). Each panel shows raw individual gastruloid profiles (light green, no y-axis normalization) and mean profiles (dark green) of three same-day replicas of SOX2 expression in immunostained gastruloids seeded, cultured, fixed, stained, and imaged in parallel on three separate plates (i.e., in each panel three same-day-replicas shown by full, dashed, and dotted lines). Each individual experiment (12 total) is composed of 25--50 gastruloids. Conditions are identical for all experiments except for experiment III in which gastruloids were mounted in PBS instead of Aqua-Poly/Mount. 
    Note that same-day replicas are significantly more reproducible (i.e., self-similar) than experiments across different days (i.e., the mean expression pattern differs more across days than across same-day replicas, something \textit{not} seen in developing embryos~\cite{Tkacik2015}).
    %
    {\bf B}:~Mean profiles as a function of relative position $x/L$ for each replica. Shaded areas are standard errors. Normalization was performed on the entire data set across all n gastruloids for a global maximum and minimum average intensity (i.e., a single max and a single mean for experiment day). Same-day replica can have absolute reproducibility (exp II--IV), where profile distributions collapse without y-axis normalization.
    {\bf C}~:~Profile variability $\sigma_{I}/\overline{I}$ as a function of relative position $x/L$ along the midline for each replica (green, line style as in A), or for the entire data set across same-day replicas (black). Panels run across four experiments as in A.  Again, same-day replicas are highly reproducible while variability profiles differ significantly across different days. 
    {\bf D}:~Positional error $\sigma_{x/L}$ calculated by error propagation from A and B for each replica. Gray lines correspond to one and two effective cell diameters $d_c$, respectively. The corresponding values in $\sigma_{x/L}$ are different between different experiments because of experiment-to-experiment variability in length (Fig.~S2F). Boundary precision is maintained near  1--2 cell diameters across all replicas (i.e., same-day and across days). 
    {\bf E}:~Variance decomposition for the SOX2 profile in experiments I and III (Methods). Plain lines correspond to the inter-plate part of the variance (for three same-day replicas) and the dashed lines to the intra-plate part of the variance. The inter-plate and intra-plate variance are represented as a fraction of the total variance of the whole population of same-day gastruloids (black lines in C). The decomposition is done in three ways: 1) on the raw profiles (black lines), 2) on normalized profiles (all profiles of individual replica are normalized by the same values, such as minimum/maximum expression levels of each replica's mean profile are set to 0/1, respectively; gray lines), and 3) on $\chi^2$-minimized profiles (all profiles of individual replicas are normalized by the same values, obtained by $\chi^2$-minimization of the mean profiles; light gray lines). Experiment I is an example of relative but not absolute reproducibility; experiment III is reproducible in absolute units, demonstrating that in principle the system is capable to generate absolute molarities of a gene product at well-defined positions along the gastruloid midline.
    {\bf F}:~Weighted average of inter-plate part of the variance, in the four experiments for either raw data, min/max normalized data, or data normalized using $\chi^2$ minimization. Internal replicas regularly achieve absolute reproducibility (i.e., no normalization, raw data comparison) better than 5\% of the total variance in the data.
    \end{justify}
\end{figure}


\clearpage
\begin{figure}
    \centering
    \includegraphics[width=\linewidth]{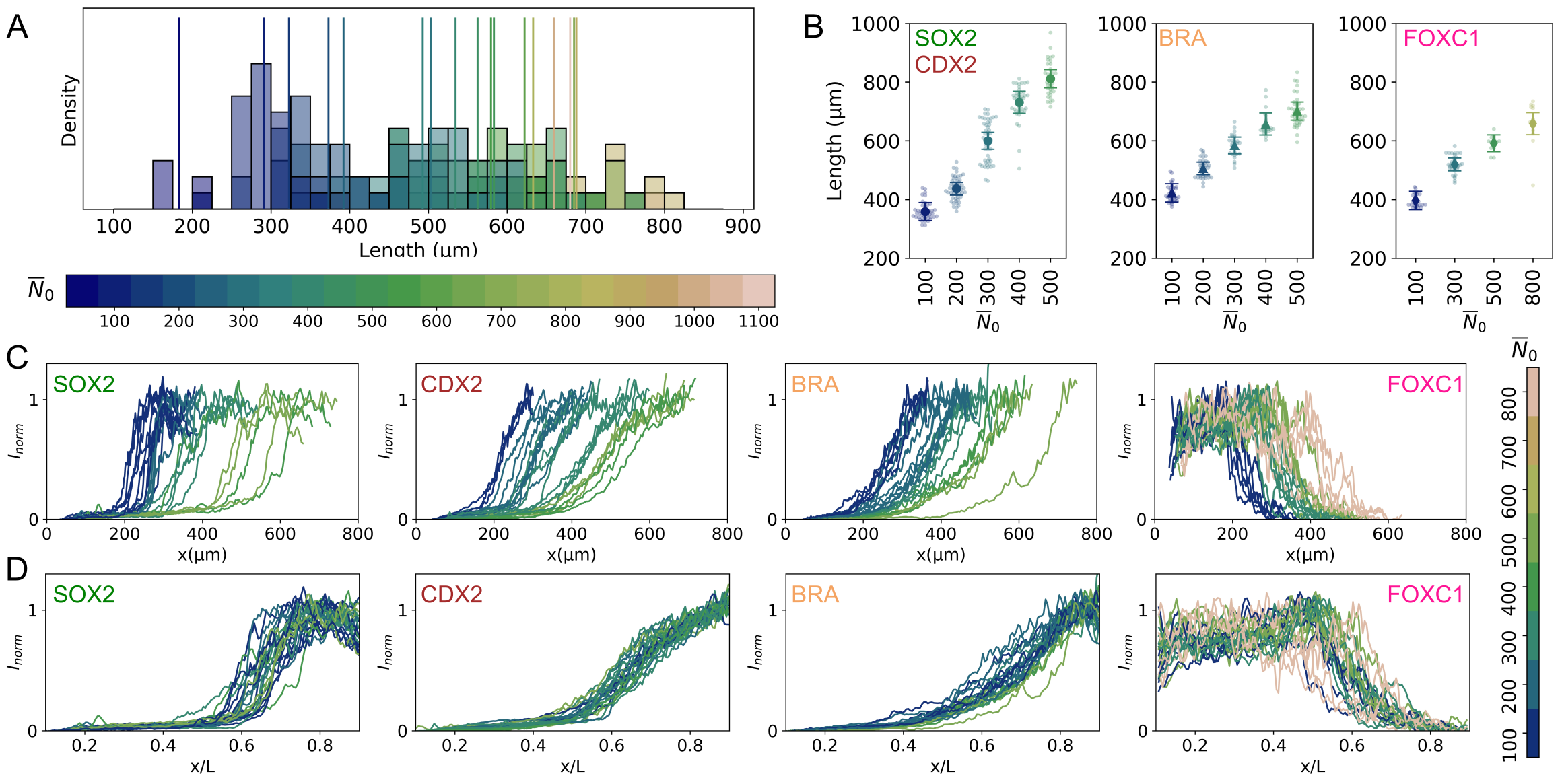}
    \caption[Scaling of gene expression in gastruloids.]{Scaling of gene expression in gastruloids.
    {\bf A}:~Midline length distribution for gastruloids at $120\,\rm{h}$ seeded with $\overline{N}_0$ ranging from 50 to 1100 (from Fig.~1B, on average 15 gastruloids per $\overline{N}_0$) with a 5.3-fold total length range. A 22-fold range in $\overline{N}_0$ results in gastruloids with a 3.8-fold range in average length $\overline{L}$ (bold vertical lines).
    {\bf B}:~Length distributions of gastruloid sets in Fig.~4 as a function of $\overline{N}_0$ (light points are individual gastruloids; dark points are average length and standard deviation per set and per gene; color code as in A). The span in length differs between experiments. For the data corresponding to SOX2 and CDX2, the 5-fold range in $\overline{N}_0$ achieves a 2.3-fold range in gastruloid length at $120\,\rm{h}$. For the data corresponding to BRA and FOXC1, a 5-fold and 8-fold range in $\overline{N}_0$ achieve a 1.7-fold range in length, respectively. 
    {\bf C}:~Individual gene expression profiles (normalized between 0 and 1 for each gastruloid individually using $I_{min}$ and $I_{max}$ as in Fig.~S6A) for each $N_0$ (color code on right) and each gene as a function of absolute position along each gastruloid's midline.
    {\bf D}:~Individual gene expression profiles (normalized as in C) for each $N_0$ (color code on right) and each gene as a function of relative position ($x/L$) along the midline.
    }
    \label{figs:scaling}
\end{figure}


\clearpage
\begin{figure}
    \centering
    \includegraphics[width=\linewidth]{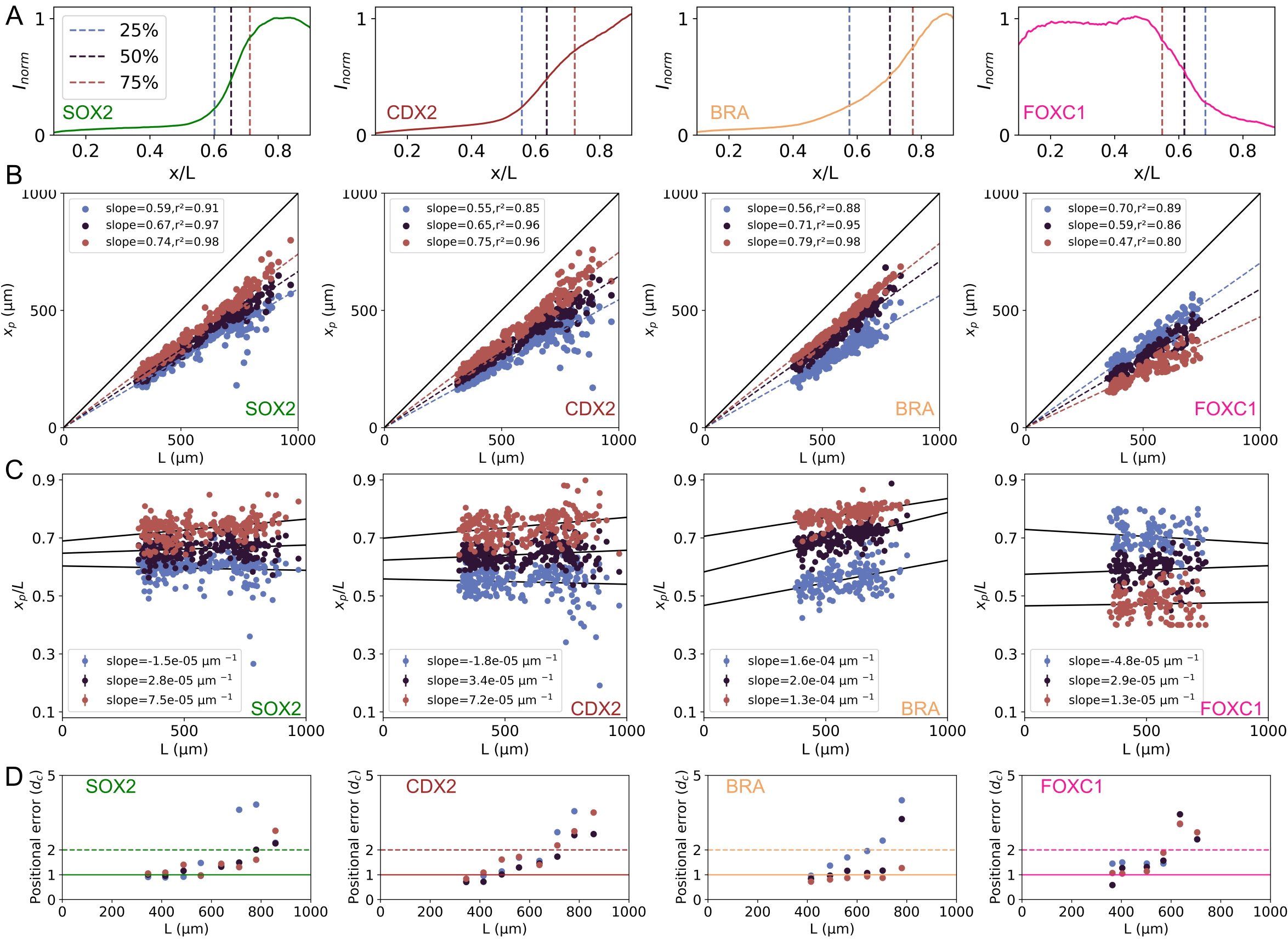}
    \caption[Limits of precision in scaled gene expression profiles]{Limits of precision in scaled gene expression profiles.
    {\bf A}:~Normalized mean expression profiles for each gene (SOX2, CDX2, BRA and FOXC1) of all gastruloids of different $N_0$. For each gene positional markers are defined at three positions corresponding to the 25\% ($x_{25}$, blue), 50\% ($x_{50}$, black), and 75\% ($x_{75}$, red) of maximum profile intensity levels (vertical dashed lines), respectively.
    {\bf B}:~Absolute positions of the 25\%, 50\% and 75\% maximum intensity levels for each gastruloid (same as in Fig.~4C) as a function of gastruloid length (same color code as above). Perfect scaling would imply $R^{2}=1$, meaning that 100\% of the observed boundary position variance is related to gastruloid length. Slope values correspond to the average position of the three positional markers in relative units $x_p/L$.
    {\bf C}: Relative position of the 25\%, 50\% and 75\% maximum intensity levels as a function of L for each gastruloid (same color code as above). Perfect scaling predicts statistical independence of the relative boundary position (50\% maximum intensity position) and the absolute gastruloid length. We performed a linear regression and found that the slopes are statistically different from zero ($p$-value$\,\le0.01$), with a 99\% confidence interval; see slopes in legend. A slope of $10^{-5}\,$µ$\rm{m^{-1}}$ means that a decrease or an increase of $300\,$µm around the case $\overline{N}_0=300$ leads to a shift of the positional marker of $\sim$1\% along the AP midline, i.e. $\sim$$6\,$µm ($\le 1 d_c$). A slope of $10^{-4}\,$µ$\rm{m^{-1}}$ (as is the case for BRA) means that a decrease or an increase of $300\,$µm leads to a shift of the positional marker of $\sim$10\% along the AP midline, i.e. $\sim$$60\,$µm ($\sim$$4 d_c$).
    For $x_{50}/L$, the slopes for the four genes SOX2, CDX2, BRA, and FOXC1 are $2.8\pm1~10^{-5}$, $3.4\pm1~10^{-5}$, $2.0\pm2~10^{-5}$ and $2.9\pm3~10^{-5}\,$µ$\rm{m^{-1}}$, respectively.
    {\bf D}:~Positional error for the three markers (same color code as above) converted in cell diameter units ($d_c$) as a function of average gastruloid length for the four genes SOX2, CDX2, BRA, and FOXC1. The range of gastruloid lengths is binned; each data point corresponds to the bin average. The positional error remains between 1--2 cells for all genes and all markers within a certain length range (up to $600\,$µm for FOXC1, up to $800\,$µm for the other genes). This range corresponds to the mean length of gastruloids in a range $100\le\overline{N}_0\le500$ for each experiment (Fig.~S10B).
    }
    \label{figs:scalinglimits}
\end{figure}


\clearpage
\section{Supplemental tables}

\begin{table}[!ht]
\centering
\begin{tabular}{|c|c|c|}
\hline
\hspace{0.5cm}\textbf{Time}\hspace{0.5cm} &
\hspace{0.5cm}\textbf{Regression slope}\hspace{0.5cm} &
\hspace{0.5cm}\textbf{Standard error}\hspace{0.5cm} 
\\ \hline
24h  & 1.002            & 0.016          \\ \hline
48h  & 0.999            & 0.010          \\ \hline
72h  & 1.001            & 0.013          \\ \hline
96h  & 1.001            & 0.013          \\ \hline
120h & 1.003            & 0.033          \\ \hline
\end{tabular}
\caption[Growth scaling regression results for individual gastruloids with $\overline{N}_0=300$] {Results of linear fits for the data points of individual gastruloids with $\overline{N}_0=300$ in the inset of Fig.~1D. We performed a linear fit for the cell count ${N}(t)/\overline{N}_{300}(t)$ as a function of the initial seed cell count ${N}_0/300$. We find that the slopes are equal to one within error bars, i.e. they are statistically indistinguishable from one at all time points.}
\end{table}

\begin{table}[!ht]
\centering
\begin{tabular}{|c|c|c|c|c|}
\hline
\hspace{0.5cm}\textbf{Antibodies}\hspace{0.5cm} & \hspace{0.5cm}\textbf{Species}\hspace{0.5cm} & \hspace{0.5cm}\textbf{Reference}\hspace{0.5cm} & \hspace{0.5cm}\textbf{Provider}\hspace{0.5cm} & \hspace{0.5cm}\textbf{Dilution}\hspace{0.5cm} \\ \hline
SOX2                & Rat              & 14-9811-80         & eBioscience       & 1:200             \\ \hline
CDX2                & Rabbit           & EPR2764Y           & Invitrogen        & 1:200             \\ \hline
BRA                 & Rabbit           & ab209665           & Abcam             & 1:150             \\ \hline
FOXC1               & Rabbit           & ab223850           & Abcam             & 1:500             \\ \hline
Anti-Rat AF488      & Donkey           &   A-21208        &     Invitrogen     & 1:500             \\ \hline
Anti-Rabbit AF647  & Donkey           &    A-31573  &   Invitrogen   & 1:500             \\ \hline
\end{tabular}
\caption[List of antibodies used for immunofluorescence staining] {List of antibodies used for immunofluorescence staining (see Methods).}

\end{table}

\begin{table}[!ht]
\centering
\begin{tabular}{|c|c|c|c|}
\hline
\textbf{Initial mESC seed size}& \hspace{0.8cm}\textbf{Midline length}\hspace{0.8cm} & \hspace{1.2cm}\textbf{Volume}\hspace{1.2cm}  & \hspace{1cm}\textbf{Cell count}\hspace{1cm} \\ \hline
\textbf{$\overline{N}_0=100$}  & $322 \pm 30\,$µm     & $4.2 \pm 0.6$ $10^6$~µ$\rm{m^3}$   & $3.0 \pm 0.4$ $10^3\,$cells      \\
                               & $(9.4\%,\ n=23)$            & $(14.5\%,\ n=23)$ & $(12.6\%,\ n=23)$    \\ \hline
\textbf{$\overline{N}_0=300$}  & $494 \pm 52\,$µm     & $1.4 \pm 0.2$ $10^7$~µ$\rm{m^3}$    & $1.0 \pm 0.1$ $10^4\,$cells       \\
                               & $(10.5\%,\ n=22)$          & $(13.6\%,\ n=22)$ & $(9.5\%,\ n=22)$     \\ \hline
\textbf{$\overline{N}_0=500$}  & $673 \pm 68\,$µm    & $2.5 \pm 0.4$ $10^7$~µ$\rm{m^3}$   & $1.9 \pm 0.2$ $10^4\,$cells      \\
                               & $(10.2\%,\ n=20)$          & $(14.6\%,\ n=20)$ & $(9.4\%,\ n=20)$     \\ \hline
\textbf{$\overline{N}_0=800$}  & $754 \pm 59\,$µm    & $3.7 \pm 0.7$ $10^7$~µ$\rm{m^3}$    & $2.7 \pm 0.4$ $10^4\,$cells       \\
                               & $(7.8\%,\ n=14)$           & $(19.3\%,\ n=14)$ & $(13.7\%,\ n=14)$    \\ \hline
\end{tabular}

\caption[Gastruloid midline length, volume and cell count at $120\,\rm{h}$ for different values of $\overline{N}_0$] {Gastruloid midline length, volume and cell count at $120\,\rm{h}$ for different values of $\overline{N}_0$. These values were obtained from the chemical dissociation of gastruloids seeded using FACS (Fig.~S1E and S7C). All numbers are means $\pm$ standard deviations across $n$ gastruloids. The coefficient of variation, defined by the ratio of standard deviation over mean, is reported in \%.}

\end{table}

\end{document}